\documentclass[fleqn,usenatbib]{mnras}

\usepackage[T1]{fontenc}

\DeclareRobustCommand{\VAN}[3]{#2}
\let\VANthebibliography\thebibliography
\def\thebibliography{\DeclareRobustCommand{\VAN}[3]{##3}\VANthebibliography}

\usepackage{graphicx}	
\usepackage{amsmath}	
\usepackage{amssymb}	
\usepackage{mathtools}

\usepackage{caption}
\usepackage{subcaption}
\usepackage{soul,xcolor}
\usepackage{xcolor}
\usepackage{hyperref}
\setstcolor{red}
\usepackage{url}

\usepackage{newtxtext,newtxmath}

\title[Magnetoacoustic waves in an inhomogeneous cylinder]{II. The effect of axisymmetric and spatially varying equilibria and flow on MHD wave modes: Cylindrical geometry}

\author[S. J. Skirvin et al.]{
S. J. Skirvin,$^{1}$\thanks{E-mail: sjskirvin1@sheffield.ac.uk}
V. Fedun,$^{1}$
Suzana S. A. Silva,$^{1}$ 
G. Verth,$^{2}$
\\
$^{1}$Plasma Dynamics Group, Department of Automatic Control \& Systems Engineering, The University of Sheffield, Sheffield, S3 7RH, UK\\
$^{2}$Plasma Dynamics Group, Department of Mathematics and Statistics, The University of Sheffield, Sheffield, S3 7RH, UK\\
}

\date{Accepted XXX. Received YYY; in original form ZZZ}

\pubyear{2021}

\begin{document}
\label{firstpage}
\pagerange{\pageref{firstpage}--\pageref{lastpage}}
\maketitle

\begin{abstract}

Magnetohydrodynamic (MHD) waves are routinely observed in the solar atmosphere. These waves are important in the context of solar physics as it is widely believed they can contribute to the energy budget of the solar atmosphere and are a prime candidate to contribute towards coronal heating. Realistic models of these waves are required representing observed configurations such that plasma properties can be determined more accurately which can not be measured directly. This work utilises a previously developed numerical technique to find permittable eigenvalues under different non-uniform equilibrium conditions in a Cartesian magnetic slab geometry. Here we investigate the properties of magnetoacoustic waves under non-uniform equilibria in a cylindrical geometry. Previously obtained analytical results are retrieved to emphasise the power and applicability of this numerical technique. Further case studies investigate the effect that a radially non-uniform plasma density and non-uniform plasma flow, modelled as a series of Gaussian profiles, has on the properties of different MHD waves. For all cases the dispersion diagrams are obtained and spatial eigenfunctions calculated which display the effects of the equilibrium inhomogeneity. It is shown that as the equilibrium non-uniformity is increased, the radial spatial eigenfunctions are affected and extra nodes introduced, similar to the previous investigation of a magnetic slab. Furthermore, azimuthal perturbations are increased with increasing inhomogeneity introducing vortical motions inside the waveguide. Finally, 2D and 3D representations of the velocity fields are shown which may be useful for observers for wave mode identification under realistic magnetic waveguides with ever increasing instrument resolution.

\end{abstract}




\section{Introduction}\label{Introduction}
Observations of the solar atmosphere show that it is replete with magnetohydrodynamic (MHD) waves. Over recent years, a great improvement in both ground and space based telescopes with finer spatial and temporal resolution has revealed the vast number of structures that may act as waveguides for MHD waves. There has been abundant evidence of the highly structured solar atmosphere supporting the propagation of these waves \citep[see e.g.][]{nak1999, dep2007, mor2011, keys2018}. 
The combination of observational data with analytical models allows for the common practice known as `solar atmospheric seismology'. This technique allows researchers to calculate properties of the solar atmospheric plasma, that may be difficult to determine directly from observations, by analysing the propagation of waves through the medium. However, for this to be a useful technique, there is a need for a high degree of accuracy in both the observations and the theoretical models. Whilst great improvement has been made on the observational side, there is still a large amount unknown about modelling wave propagation in general realistic solar waveguides.

Analytical theory of MHD waves with an application to solar physics is widely accepted as being introduced many decades ago \citep{Hain1958, Spr1979, wilson1979, Wilson1980, rob1981I, rob1981II, Spruit1982, edrob1982, edrob1983}. These early works provided a description of the properties of MHD waves in both Cartesian and cylindrical geometries of the waveguide comprised of different magnetic interfaces. Since then, adaptations and specific case studies have been undertaken to explore the properties of MHD waves under more complicated yet realistic plasma configurations. A non exhaustive list includes investigating waves in non-uniform magnetic slabs \citep{arr2007, lopin2015a, lopin2015b, li2018, Skirvin2021}, curved magnetic slabs \citep{Ver2006}, twisted magnetic cylinders \citep{ErdFed2007, ErdFed2010, ter2012, ter2018} and magnetic cylinders with vortex flows \citep{cher2017,Cher2018, tsap2020}. It is widely known that within a non-uniform plasma in ideal MHD, waves propagating at specific frequencies may resonate with the local plasma. This occurs in continuum regions where the wave propagates at either the local Alfv\'{e}n speed or the local tube (cusp) speed. At these locations, it is to be expected that wave energy can be extracted by processes such as resonant absorption \citep{goo1995, kepp1996, goo2011}. Undertaking an investigation into wave damping mechanisms relies on treating the wave frequency as a complex quantity, where the imaginary component provides information on either wave damping or any wave related instabilities. This phenomenon is not considered in this work however relevant studies can be found in e.g. \citep{hey1983, Yu2021}. The present study includes plasma (gas) pressure in the analysis however avoids resonantly damped modes by assuming that the wave frequency is purely real. It should be noted that previous works have investigated the resonantly damped modes however, in their analysis assume that the plasma-$\beta$ is zero, which ultimately removes the slow magnetoacoustic modes from their analysis \citep{VanDoor2004, soler2013, soler2017ApJ...850..114S, soler2019}.

It is well known that the solar atmosphere and features within it are highly non-uniform, mainly due to the fine magnetic fields which permeate throughout \citep[e.g.][]{Will2020}. This non-uniformity has an important affect on the propagation and observation of MHD waves. It was shown in \citet{Skirvin2021} (hereafter Paper 1) that an inhomogeneous plasma equilibrium changes the eigenvalues for trapped wave modes dependant upon the scale of the inhomogeneity. If the plasma is highly non-uniform then the permittable bands within which MHD waves can propagate become narrower, whereas the continuum regions, where physical damping processes can take place, spread a wider range of phase speeds. The non-uniform equilibrium can also affect the spatial distributions of observable eigenfunctions produced by MHD wave propagation. It was found that in coronal slab structures, slow body modes are more affected by large inhomogeneities in density over the width of a magnetic slab, the same is true for body modes in a photospheric slab. 

Furthermore, the effect of non-uniformity has important implications for identifying wave modes in observations. It is well known that the theoretical Alfv\'{e}n mode along with the slow and fast magnetoacoustic modes only exist in pure form in a uniform plasma of an infinite extent. In such a scenario the Alfv\'{e}n wave propagates as a purely incompressible vortical perturbation with only magnetic tension acting as the restoring force. Furthermore, magnetoacoustic waves propagate as compressible disturbances which can be identified through observations of plasma intensity perturbations with a combination of magnetic tension and total pressure acting as the restoring forces. In reality however, the solar atmosphere is highly non-uniform and not infinite. Consequently it is difficult to interpret observations of MHD wave modes as one of the three distinct wave modes in a uniform plasma. In a non-uniform plasma MHD waves have mixed properties and cannot be classified as pure Alfv\'{e}n or pure magnetoacoustic \citep{goo2019}.

It has been shown analytically that even a simple discontinuity in plasma equilibrium such as a piece-wise true discontinuous density (similar to that modelled as a magnetic cylinder) that the fundamental radial non-axisymmetric magnetoacoustic mode (kink mode) should in fact be interpreted as a surface Alfv\'{e}n wave \citep{Wen1979, goo2012}. The analytical study conducted by \citet{goo2012} focused on the role that vorticity plays when the plasma in non-uniform. For a pure Alfv\'{e}n wave the displacements are vortical everywhere however for a pure magnetoacoustic wave there is zero vorticity. As the piece-wise discontinuity is replaced with a continuous profile, vorticity is spread out over the whole interval covered by the Alfv\'{e}n continuum, where the density is inhomogeneous. The role of vorticity in a non-uniform magnetic flux tube is further investigated in this work for both a non-uniform equilibrium density and a non-uniform background plasma flow.

The effect of a steady background plasma flow on the properties of MHD waves has been previously studied in a cylindrical geometry. An investigation by \citet{terra2003} derived and solved the dispersion relation for MHD waves in a uniform magnetic cylinder with a uniform background plasma flow. Further studies have looked at the potential a steady flow may have for the onset of the Kelvin-Helmholtz instability, which may provide a turbulent cascade of energy that could heat localised plasma \citep{zhel2012, zhe2013}. However, little research has been conducted which investigates the effect a non-steady plasma flow may have on the properties of MHD waves in a magnetic cylinder due to the difficulty of analytically deriving a dispersion relation.

In this work, the approach to that described in Paper 1 is slightly modified with the main difference in the coordinate system used which affects the vector operators in the analytical investigation. In this investigation, the numerical technique developed previously is applied to initially uniform cylindrical waveguides to reproduce previously obtained analytical results of a uniform cylinder, also including a steady background flow. Thereafter, the internal spatial profile of plasma density and background flow is allowed to be non-uniform in the shape of a series of Gaussian profiles, which cannot be investigated purely analytically, with discussions about the observable differences in wave properties due to this non-uniform equilibria.

This paper is organised as follows: the ideal MHD equations describing motions in a radially non-uniform cylinder are presented in Section \ref{Method}. In Section \ref{Applications} the numerical tool is applied to previously studied cases with known analytical results, the analytical and numerical results are compared. Further investigations of non-uniform density cases which cannot be studied analytically are discussed in Section \ref{gaussian_density}. The MHD wave behaviour in a uniform coronal cylinder with a non-uniform background plasma flow is analysed in Section \ref{gaussian_flow}. Lastly a summary and discussion of the results obtained in this paper can be found in Section \ref{conclusions}.

\section{Method}\label{Method}

This work adopts a cylindrical geometry in the form $(r, \varphi, z)$. The initial equilibrium is allowed to be radially spatially dependant for all variables and has background magnetic field vector components in the form $(0, 0, B_{0z})$ and background velocity field vector components $(0, 0, U_{0z}(r))$. In this work $U_{0z}$ is taken to be positive which corresponds to a flow in the positive vertical direction. Since the equilibrium quantities depend on $r$ only, the perturbed quantities can be Fourier-analysed with respect to the ignorable coordinates $\varphi, z$ and time $t$ and put proportional to: 
\begin{equation*}
\text{exp}\left[i\left(m\varphi + kz - \omega t\right)\right],
\end{equation*}
where $m$ is the azimuthal wave number, $k$ the vertical wavenumber and $\omega$ the wave frequency.

The system of linearised MHD equations (see Paper 1) can be reduced to a system of two differential equations containing the total pressure perturbation $\hat{P}_T$ and the radial displacement perturbation $r\hat{\xi}_r$ as previously done in \cite[see e.g.][]{sak1991, goo1992, ErdFed2007, ErdFed2010}. It should be noted that the full set of equations written in these works listed above consider an equilibrium that includes background magnetic and velocity azimuthal components. In this work these physical properties are ignored and as a result the governing differential equations can be written as:

\begin{equation}\label{rxi_r_diff}
    D\frac{d}{dr}\left(r\hat{\xi}_r\right)= - C_1 r \hat{P}_T,
\end{equation}
\begin{equation}\label{P_diff}
    D\frac{d\hat{P}_T}{dr}=C_2 \hat{\xi}_r,
\end{equation}
where,
\begin{equation}\label{D}
    D = \rho\left(c^2+v_A^2 \right)\left(\Omega^2-\omega_A^2\right)\left(\Omega^2-\omega_c^2\right),
\end{equation}
\begin{equation}\label{Omega}
\Omega = \omega - k U_{0z}(r),
\end{equation}
\begin{equation}
c^2 = \frac{\gamma P}{\rho},\ \ \ \ v_A^2 = \frac{B_{0z}^2}{\mu\rho}, \ \ \ \ \omega_A^2 = \frac{k^2B_{0z}^2}{\mu\rho}, \ \ \ \ \omega_c^2 = \frac{\omega_A^2 c^2}{\left(c^2 + v_A^2\right)},
\end{equation}
\begin{equation}\label{C_1}
    C_1 = \Omega^4 - \left(c^2+v_A^2\right)\left(\frac{m^2}{r^2}+k^2\right)\left(\Omega^2-\omega_c^2\right),
\end{equation}
\begin{equation}\label{C_2}
    C_2 = D\rho\left(\Omega^2-\omega_A^2 \right). 
\end{equation}

Variables $\rho$, $P$, $B_{0z}$, $\gamma$ and $\mu$ denote plasma density, plasma pressure, vertical magnetic field, ratio of specific heats (taken $\gamma = 5/3$) and the magnetic permeability respectively. Quantities $c^2$, $v_A^2$, $\omega_A^2$ and $\omega_c^2$ define the squares of the local sound speed, Alfv\'{e}n speed, Alfv\'{e}n frequency and cusp frequency respectively. Equations (\ref{rxi_r_diff})-(\ref{C_2}) are the simplified set of equations from earlier studies which can be retrieved by setting the azimuthal components $v_{\varphi} = B_{\varphi} = 0$ \cite[see e.g.][]{sak1991,goo1992}. Equation (\ref{Omega}) describes the Doppler shifted frequency due to the presence of the background plasma flow. It is clear that with no field alligned flow present in the model that this equation simply reduces to the wave frequency. Furthermore, it can be seen that if the plasma flow is steady (spatially uniform) then this expression describes the steady Doppler shift shown previously in \citet{nak1995, terra2003}. However, this quantity is now a function of radially variable $r$ and as a result the effect of the Doppler shifted frequency depends on the local background plasma flow at that location. The set of Equations (\ref{rxi_r_diff})-(\ref{C_2}) provide the full equations for any cylindrical equilibrium with a radially varying field aligned flow. It should also be noted that they describe any cylindrical equilibrium which is non-uniform in the direction of spatial coordinate $r$ (e.g. $\rho(r)$), as a result all quantities would also depend on $r$ in such an equilibrium. Equations (\ref{rxi_r_diff})-(\ref{P_diff}) can be combined to create a single differential equation in either $r\hat{\xi}_r$:
\begin{equation}\label{rxi_r_diff_eqn}
    \frac{d}{dr}\left[ f(r)\frac{d}{dr}\left(r\hat{\xi}_r\right)\right] - g(r)\left(r\hat{\xi}_r\right) = 0,
\end{equation}
where,
\begin{equation}
f(r) = \frac{D}{rC_1},
\end{equation}
\begin{equation}
g(r) = - \frac{C_2}{rD},
\end{equation}
or $\hat{P}_T$:
\begin{equation}\label{P_diff_eqn}
    \frac{d}{dr}\left[ \Tilde{f}(r)\frac{d\hat{P}_T}{dr}\right] - \Tilde{g}(r)\hat{P}_T = 0,
\end{equation}
where,
\begin{equation}
\Tilde{f}(r) = \frac{rD}{C_2},
\end{equation}
\begin{equation}
\Tilde{g}(r) = - \frac{rC_1}{D}.
\end{equation}

Similar to the scenario encountered in Paper 1, both Equations (\ref{rxi_r_diff_eqn}) and (\ref{P_diff_eqn}) have no known closed form analytical solutions for arbitrary profiles of plasma equilibrium properties such as density, magnetic field, background flow etc. Therefore, investigating the properties of wave modes propagating within an equilibrium which is non-uniform must be done numerically. 

The numerical algorithm employed in this work is the cylindrical counterpart of the approach presented in Paper 1. Trapped modes are investigated and complex frequencies are ignored such that the waves are assumed to be evanescent away from the cylinder boundary located at $r=a$. The numerical shooting method is implemented to solve Equations (\ref{rxi_r_diff_eqn}) and (\ref{P_diff_eqn}) ensuring continuity of $\hat{P}_T$ and $\hat{\xi}_r$ across the boundary. The application of this technique is modified slightly in this work compared to the magnetic slab counterpart which investigated waves in a Cartesian coordinate system. It can be seen that Equations (\ref{rxi_r_diff_eqn}) and (\ref{P_diff_eqn}) become singular at $r=0$, as a result the numerical shooting method will fail here. To overcome this, the boundary conditions in this work also require to be matched at the (numerical) centre of the cylinder ($r\ll a$) - dependant upon the wave mode being investigated. For example for the kink mode, $\hat{P}_T$ must be continuous across the boundary at $r=a$ but must also be equal to zero at the centre of the cylinder. Likewise for the sausage mode, which does not perturb the waveguide axis, it is required that $\hat{\xi}_r$ must be equal to zero at the center of the cylinder and continuous across $r=a$. Waves which have frequencies and wavenumbers that satisfy these conditions whilst also being solutions to the governing differential equations will be classed as eigenvalues of the system and can be plotted on the dispersion diagram.

For later reference it may be important to note that the component of the displacement vector of magnetic surfaces perpendicular to the magnetic field lines $\hat{\xi}_{\varphi}$ can be related to $\hat{P}_T$ and general plasma properties by:

\begin{equation}\label{xi_perp}
    \left(\Omega^2 - \omega_A^2\right)\hat{\xi}_{\varphi} = \frac{i}{\rho B_{0z}}\left(g_B \hat{P}_T\right),
\end{equation}

where, 
\begin{equation*}
    g_B= \mathbf{\left(k \times B\right)}_r = \frac{m}{r}B_{0z}.
\end{equation*}

For the case of a uniform cylinder in the absense of background inhomogeneity in terms of plasma, magnetic field or plasma velocity, Equation (\ref{xi_perp}) reduces to:
\begin{equation}\label{azimuthal_comp}
    \hat{\xi}_{\varphi} = \frac{i}{\rho (\omega^2 - k^2v_A^2)}\frac{m}{r}\hat{P}_T,
\end{equation}
which is a previously obtained analytical result for a uniform/non-uniform static magnetic cylinder with a straight background magnetic field \citep{goo1992,goe2004, goo2009, Rud2009, Priest2014}.

For a non-uniform plasma, including a non-uniform background plasma flow, the variable $\Omega$ and the characteristic frequencies $\omega_A$, $\omega_c$ will depend on the spatial variable $r$, under these conditions resonant absorption is rather the rule than the exception. The regions bounded by:
\begin{equation}\label{alfven_flow_continuum}
    \omega = \omega_f (r) \pm \omega_A (r),
\end{equation}
\begin{equation}\label{cusp_flow_continuum}
    \omega = \omega_f (r) \pm \omega_c (r),
\end{equation}
define the Doppler shifted continua where wave modes can be resonantly damped, corresponding to singularities in the governing Equations (\ref{rxi_r_diff})-(\ref{P_diff}). Within these regions the wave frequency becomes a complex quantity, as a result these are the regions which are not considered in this work.

\section{Comparison with known solutions}\label{Applications}

   \begin{figure*}
   \centering
   \begin{subfigure}{.49\textwidth}
        \centering
        \includegraphics[width=9.cm]{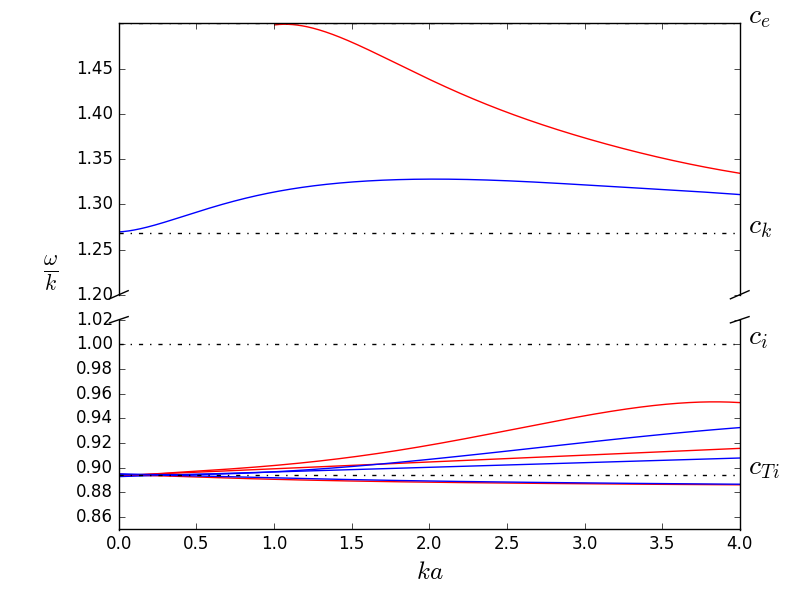}
        \caption{}
        \label{photospheric_uniform}
    \end{subfigure}
   \begin{subfigure}{.49\textwidth}
        \centering
        \includegraphics[width=9.cm]{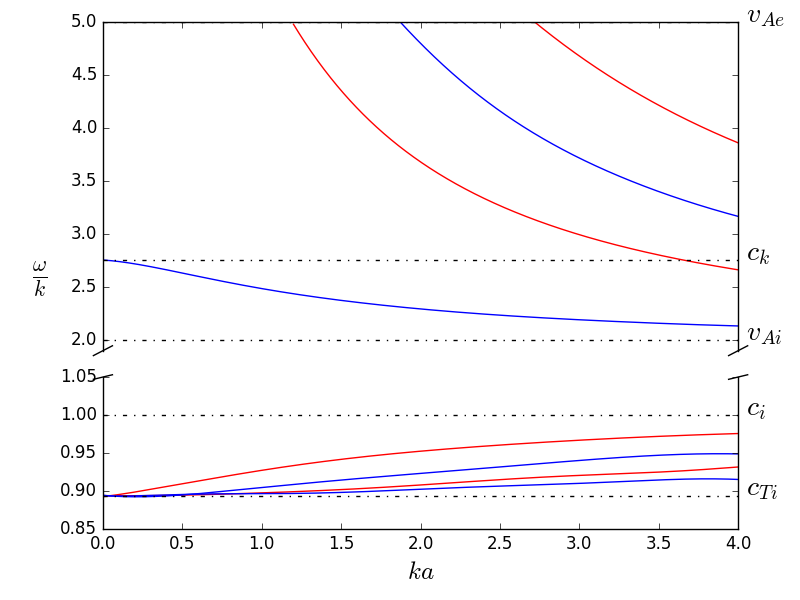}
        \caption{}
        \label{coronal_uniform}
    \end{subfigure}   

    \caption{The numerical solutions plotted on the dispersion diagram for a uniform magnetic cylinder under (a) photospheric conditions given by $c_e = 1.5c_i$, $v_{Ai} = 2c_i$ and $v_{Ae} = 0.5c_i$. (b) Coronal conditions given by $c_e = 0.5c_i$, $v_{Ai} = 2c_i$ and $v_{Ae} = 5c_i$. Red curves denote sausage mode, blue curves show kink mode. Figures replicate those shown in Figure 3 and Figure 4 in \citet{edrob1983}.}
    \label{uniform_cylinder}
   \end{figure*}

   \begin{figure*}
   \centering
   \begin{subfigure}{.49\textwidth}
        \centering
        \includegraphics[width=9.2cm]{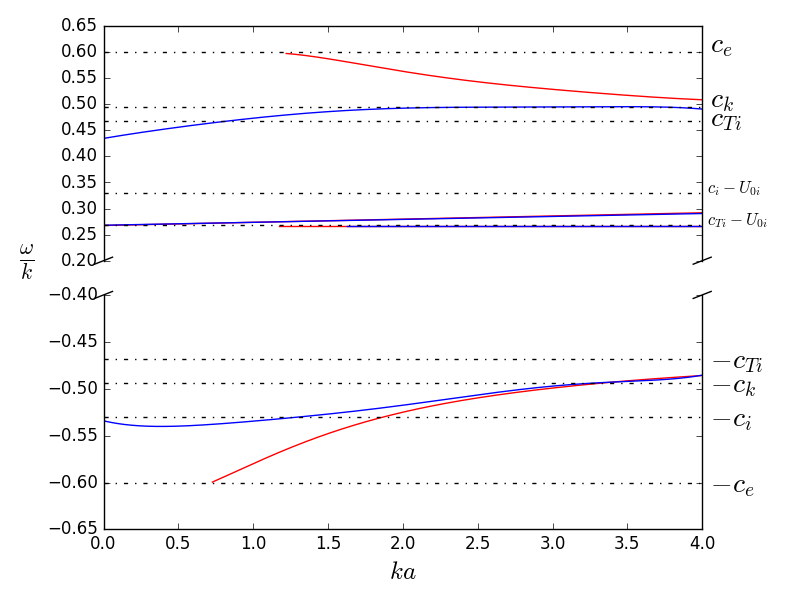}
        \caption{}
        \label{photospheric_uniform_steadyflow}
    \end{subfigure}
   \begin{subfigure}{.49\textwidth}
        \centering
        \includegraphics[width=9.2cm]{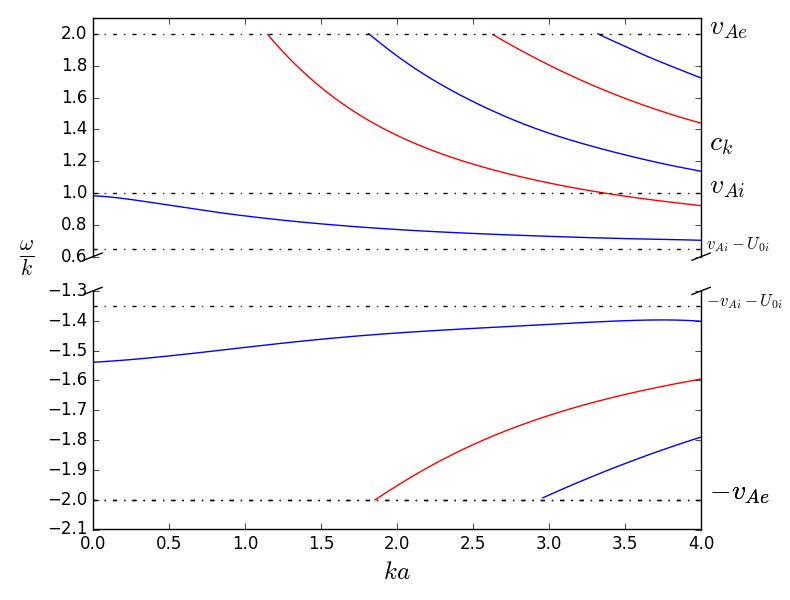}
        \caption{}
        \label{coronal_uniform_steadyflow}
    \end{subfigure}   

    \caption{The numerical solutions plotted on the dispersion diagram for a uniform magnetic cylinder with a steady background plasma flow under (a) photospheric conditions given by $c_e = 0.6v_{Ai}$, $c_{i} = 0.53v_{Ai}$, $v_{Ae} = 0$ and $U_{0i}=0.2v_{Ai}$. (b) Coronal conditions given by $c_e = 0.07v_{Ai}$, $c_{i} = 0.11v_{Ai}$, $v_{Ae} = 2v_{Ai}$ and $U_{0i}=0.35v_{Ai}$. Red curves denote sausage mode, blue curves show kink mode. Figures replicate those shown by Figure 10 and Figure 5 in \citet{terra2003} respectively.}
    \label{uniform_cylinder_steadyflow}
   \end{figure*}
In this section the numerical approach is tested against analytical results previously obtained for MHD waves in a magnetic cylinder, which may represent the majority of structures observed in the solar atmosphere that may support the propagation of such waves. In all of these cases, the analytical dispersion relation is not required and the eigenvalues are obtained purely numerically.

\subsection{Uniform magnetic cylinder}\label{uniform_cylinder_subsec}

The foundations of investigating MHD waves in a cylindrical geometry with an application to solar physics were presented in \citet{edrob1983} where the dispersion relations were derived for MHD waves in a uniform cylinder embedded in a magnetic environment under both photospheric and coronal conditions. This work was an extension of previous studies by \citet{wilson1979, Wen1979,Spruit1982} which analysed specific types of oscillations in a magnetic cylinder. Recovering the dispersion diagrams introduced in \citet{edrob1983} is the starting point for testing the numerical algorithm in a cylindrical geometry. Both photospheric and coronal dispersion diagrams are shown in Figure (\ref{uniform_cylinder}) and can be directly compared to Figure 3 and Figure 4 in \citet{edrob1983} under photospheric and coronal conditions respectively. Figure (\ref{photospheric_uniform}) shows the resulting dispersion diagram for a uniform cylinder under photospheric conditions, where the correct eigenvalues are identified that agree with the analytical results of \citet{edrob1983}. The branches of fast kink and surface sausage waves are trapped between $c_k$ and $c_e$, slow body modes are trapped between $c_{Ti}$ and $c_i$ and slow kink and sausage surface waves are seen propagating at speeds just below $c_{Ti}$. Figure (\ref{coronal_uniform}) displays the obtained dispersion diagram for a uniform cylinder under coronal conditions. Fast body modes propagate at speeds above $v_{Ai}$ and experience a cut-off at $v_{Ae}$ where at speeds faster than this they become leaky \citep{Wil1981, Sten1998, sten1999}. The fundamental kink branch can be seen trapped between $v_{Ai}$ and $c_k$ which tends towards $c_k$ in the long wavelength limit, this is in agreement with the analytical results of \citet{edrob1983}. Furthermore, in a coronal environment, slow body modes are found to be trapped between $c_{Ti}$ and $c_i$. This uniform cylinder is a very basic model of waveguides observed in the solar atmosphere, yet it is reassuring that the numerical algorithm obtains the known real eigenvalues under the new boundary conditions and geometry imposed in Section \ref{Method}.

\subsection{Magnetic cylinder with steady flow}\label{cylinder_magnetic_steadyflow}

It is well known that the addition of a bulk background plasma flow introduces new physics into the observed wave modes. \citet{nak1995} conducted an analytical study into the effect that a steady plasma flow has on magnetoacoustic waves in a magnetic slab. They found that the presence of a background flow introduces an observed Doppler shift of the wave frequency. This frequency shift alters the physics slightly as wave modes may be shifted into windows where they are not permitted to propagate as trapped modes. These results were also recovered in Paper 1 using the numerical shooting method rather than the analytical approach.   

In this section, the analytical results from a previous study \citep{terra2003}  are recovered which investigate the effect that a steady flow has on the MHD wave modes of a magnetic cylinder. The authors took a uniform magnetic cylinder model adopted from \citet{edrob1983} and incorporated a steady background plasma flow, similar to that done by \citet{nak1995} but in a cylindrical geometry. The authors came to a very similar conclusion to that of the magnetic slab with a steady flow counterpart. Namely, the inclusion of a steady background plasma flow changes the properties of magnetoacoustic waves both qualitatively and qualitatively, in the sense that the flow provides an observed Doppler shift to the wave modes which may shift the cut off values and propagation speeds in both the short and long wavelength limits. Shown in Figure (\ref{uniform_cylinder_steadyflow}) are the resulting dispersion diagrams obtained using the numerical technique for waves under photospheric and coronal conditions in a magnetic cylinder with a background steady flow. Figure \ref{photospheric_uniform_steadyflow} shows the dispersion diagram for a magnetic cylinder under typical photospheric conditions with an internal plasma flow of $U_{0i} = 0.2v_{Ai}$. This Figure is representative of Figure 10 in \citet{terra2003}. The asymmetry between forward propagating and backward propagating waves can be clearly seen by the structure and cut-off values of the fast forward and backward surface modes. Under these conditions the backward sausage and kink body modes are shifted into a region where they no longer exist as trapped modes. However, the background plasma flow is not strong enough to shift the forward body and surface modes into the leaky regime, instead these modes are shifted relative to the flow speed. Figure \ref{coronal_uniform_steadyflow} shows the dispersion diagram for magnetoacoustic waves in a coronal magnetic cylinder with an internal steady background flow of $U_{0i} = 0.35v_{Ai}$. Similar to the photospheric case, it is clear that all wave modes are shifted by a constant frequency due to the background flow. This effect can be clearly seen by the cut-off wavenumbers between the forward and backward propagating fast body modes.

The numerical approach has now been tested against two separate well known analytical results in a cylindrical geometry. Therefore, it is now appropriate to modify the initial equilibrium to a more mathematically complex scenario which cannot be fully investigated analytically.

\section{Inhomogeneous plasma density}\label{gaussian_density}

\begin{figure*}
  \includegraphics[width=\textwidth]{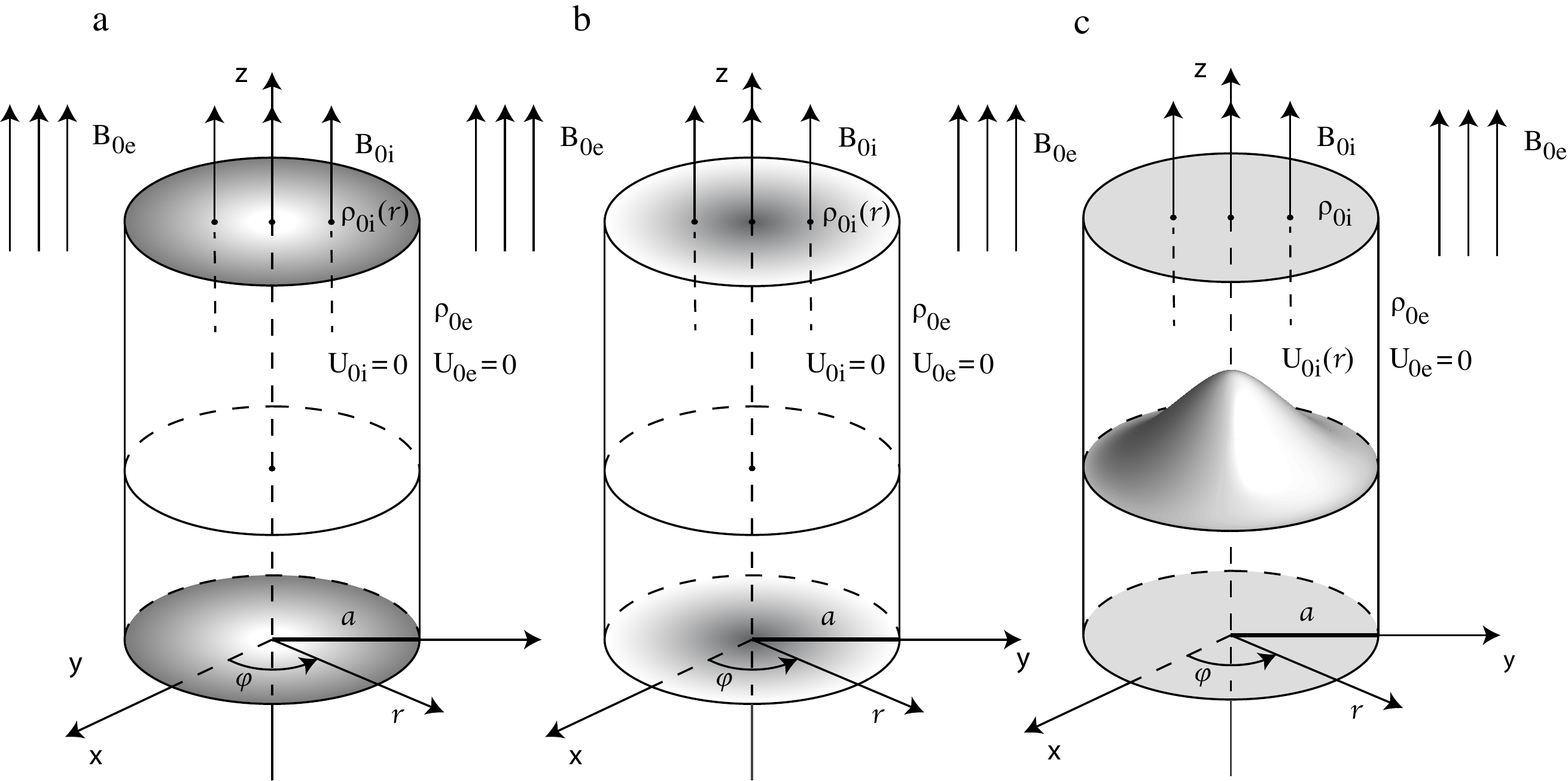}
  \caption{Cartoon depicting the equilibrium configuration of a radially dependant non-uniform magnetic flux tube in the solar atmosphere. Three separate cases are studied in this work. An inhomogeneous density magnetic flux tube under (a) photospheric  conditions and (b) coronal conditions with no plasma flow. These types of equilibria may represent features observed in the solar atmosphere e.g. sunspots and coronal loops respectively. The equilibrium density profile inside the magnetic flux tube $\rho_{0i}(r)$ is denoted by the shaded contours, with a darker shade representing a (locally) more dense plasma. The actual profiles of $\rho_{0i}(r)$ investigated in this work for the photospheric case are shown in Figure (\ref{Gaussian_density_profiles_photospheric}) and for the coronal case shown in Figure (\ref{Gaussian_density_profiles}). Finally a uniform magnetic flux tube with a non-uniform internal background plasma flow $U_{0i}(r)$ shown as 3D Gaussian shape, see panel (c), is considered, with radial profiles as demonstrated in Figure (\ref{Gaussian_flow_profiles}). This case is applicable for some jet-like structures observed in the solar atmosphere.}
  \label{cylinder_cartoon}
\end{figure*}


In this section, the equilibrium internal plasma density is considered to be inhomogeneous. For all the following case studies considered in this section, the spatially non-uniform plasma density is modelled as a series of Gaussian profiles. These profiles are modelled using the expression:
\begin{displaymath}
    \rho_i(r) = \rho_{0e} + \left(\rho_{0i} - \rho_{0e}\right)\text{exp}\left(-\frac{(r-r_0)^2}{W^2}\right),
\end{displaymath}
where $r_0$ is the centre of the Gaussian located at $r = 0$, $W$ is the standard deviation (i.e. the width) of the density distribution and $\rho_{0i}$ is the internal plasma density at $r_0$. A sketch of the non-uniform cylinder is shown in Figure (\ref{cylinder_cartoon}). Total pressure balance is achieved by a variation in equilibrium temperature to maintain a constant gas pressure across the flux tube.

\subsection{Photospheric conditions}\label{nonuniform_photospheric_density}
   \begin{figure}
   \centering
    \includegraphics[width=8.5cm]{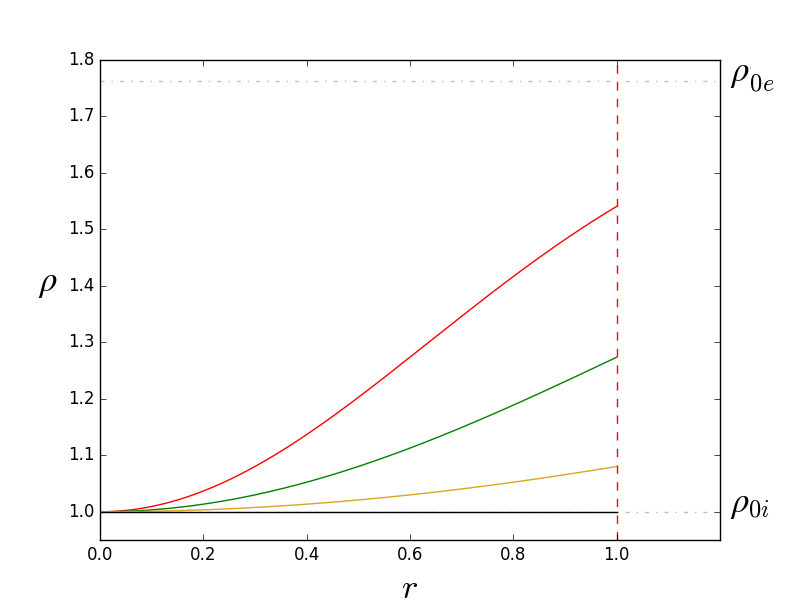}
    \caption{Gaussian background density profiles studied in this work for a non-uniform magnetic cylinder under photospheric conditions. $W = 10^5$ (black), $W = 3$ (yellow), $W = 1.5$ (green) and $W = 0.9$ (red).}
    \label{Gaussian_density_profiles_photospheric}
   \end{figure}

   \begin{figure*}
   \centering
   \begin{subfigure}{.49\textwidth}
        \centering
        \includegraphics[width=9.cm]{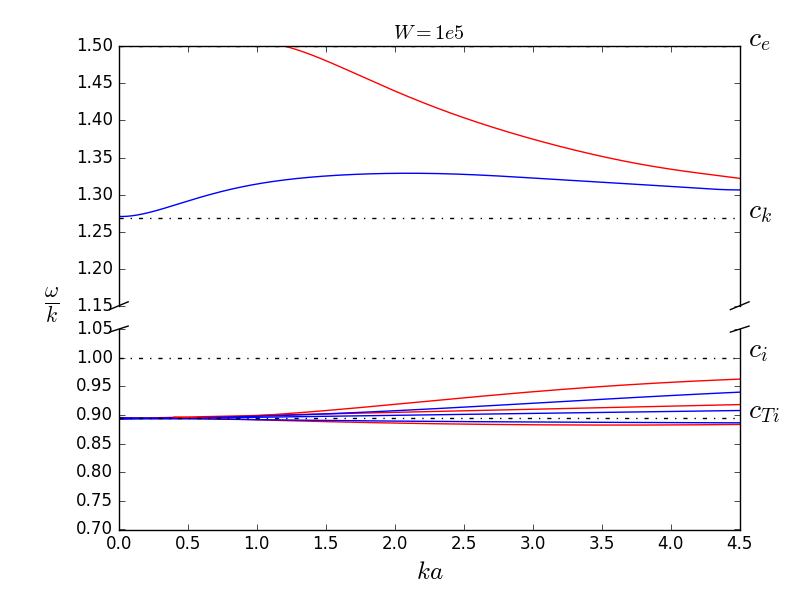}
        \caption{}
        \label{Photospheric_density_dds_1e5}
    \end{subfigure}
   \begin{subfigure}{.49\textwidth}
        \centering
        \includegraphics[width=9.cm]{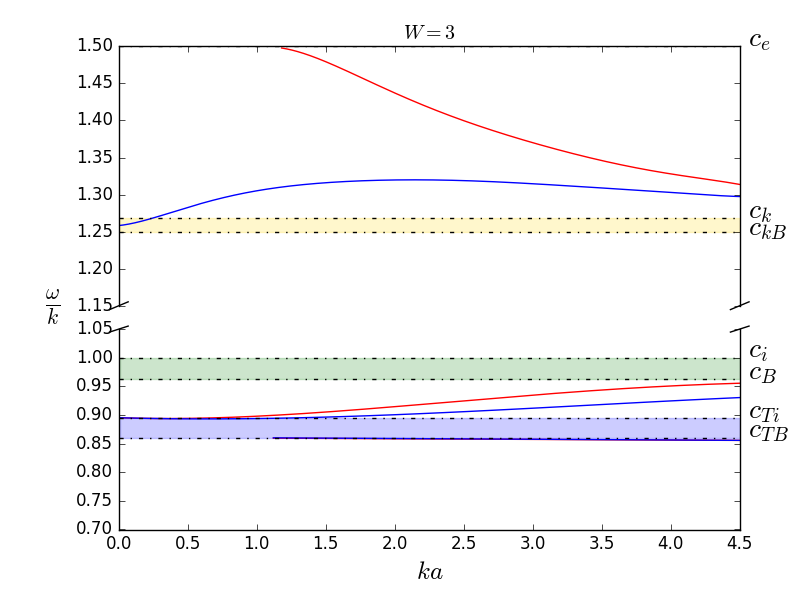}
        \caption{}
        \label{}
    \end{subfigure}   
   \begin{subfigure}{.49\textwidth}
        \centering
        \includegraphics[width=9.cm]{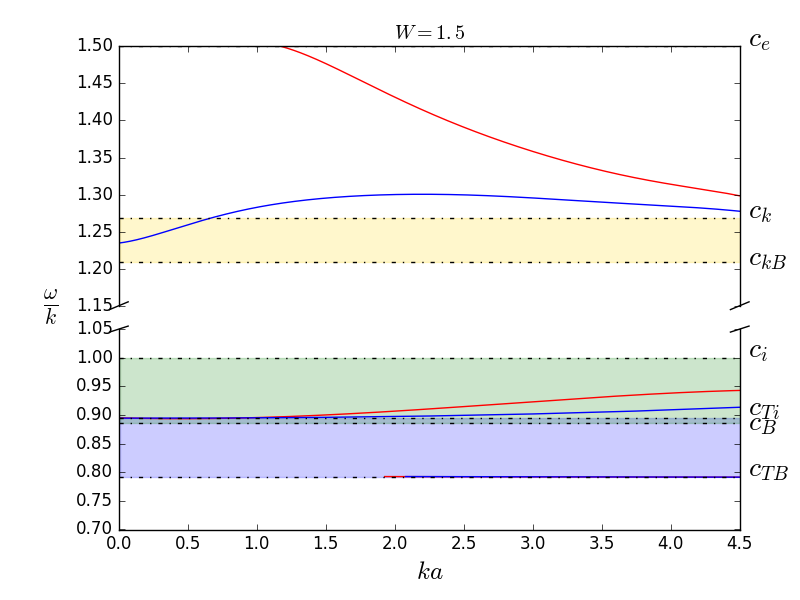}
        \caption{}
        \label{}
    \end{subfigure} 
   \begin{subfigure}{.49\textwidth}
        \centering
        \includegraphics[width=9.cm]{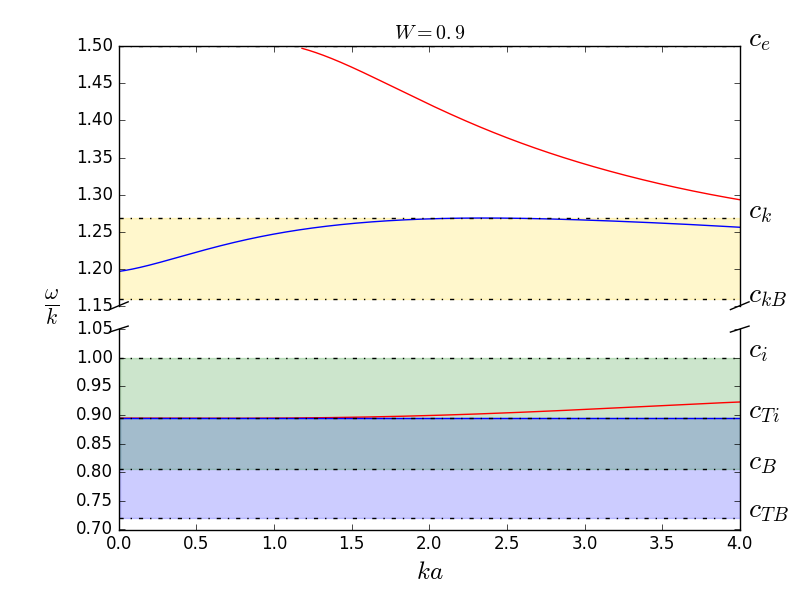}
        \caption{}
        \label{}
    \end{subfigure} 
    \caption{Dispersion diagrams for magnetoacoustic waves in a photospheric cylinder with a background plasma density in the form of Gaussian profiles shown in Figure (\ref{Gaussian_density_profiles_photospheric}). (a) $W=10^5$ corresponding to a uniform flow, (b) $W=3$, (c) $W=1.5$ and (d) $W=0.9$. Red curves denote sausage mode, blue curves show kink mode. Shaded regions represent the non-uniform bands due to the equilibrium inhomogeneity. The slow continuum (blue shaded region), inhomogeneous sound speed band (green shaded region) and inhomogeneous kink speed band (orange shaded region) are all shown.}
    \label{Photospheric_density_dds}
   \end{figure*}

   \begin{figure*}
   \centering
   \begin{subfigure}{.49\textwidth}
        \centering
        \includegraphics[width=9.cm]{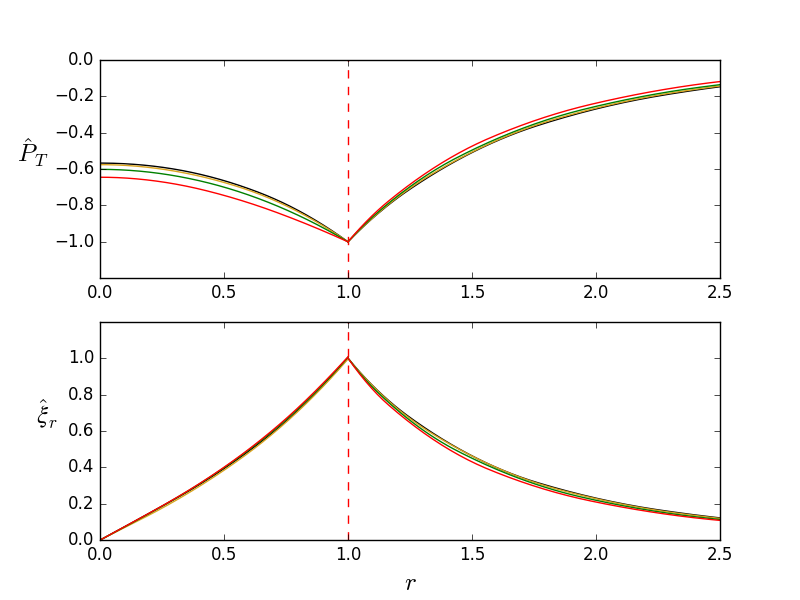}
        \caption{}
        \label{Coronal_density_dds_1e5}
    \end{subfigure}
   \begin{subfigure}{.49\textwidth}
        \centering
        \includegraphics[width=9.cm]{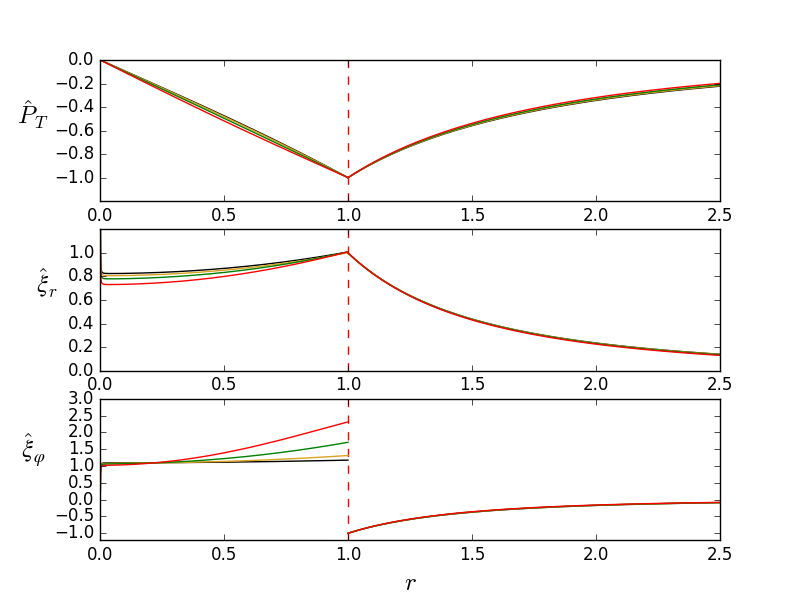}
        \caption{}
        \label{photospheric_density_fast_kink}
    \end{subfigure}   
   \begin{subfigure}{.49\textwidth}
        \centering
        \includegraphics[width=9.cm]{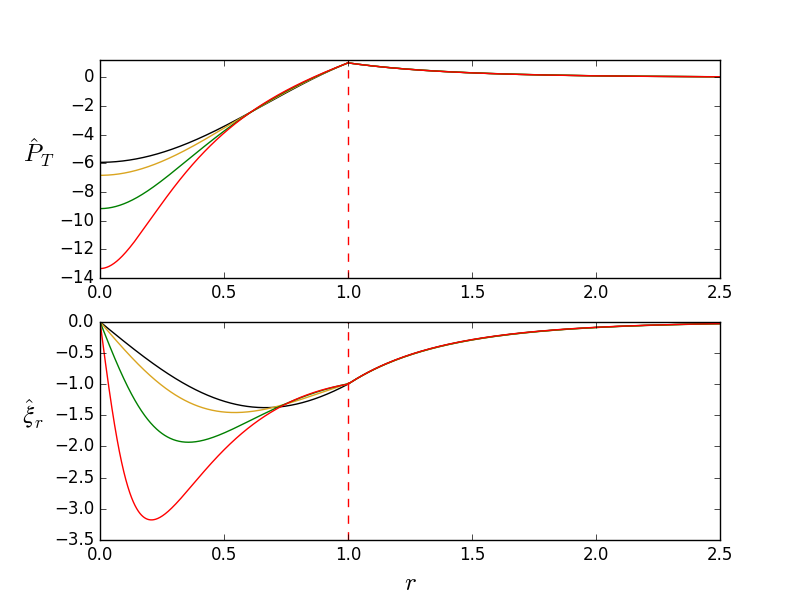}
        \caption{}
        \label{slow_body_sausage_photospheric}
    \end{subfigure} 
   \begin{subfigure}{.49\textwidth}
        \centering
        \includegraphics[width=9.cm]{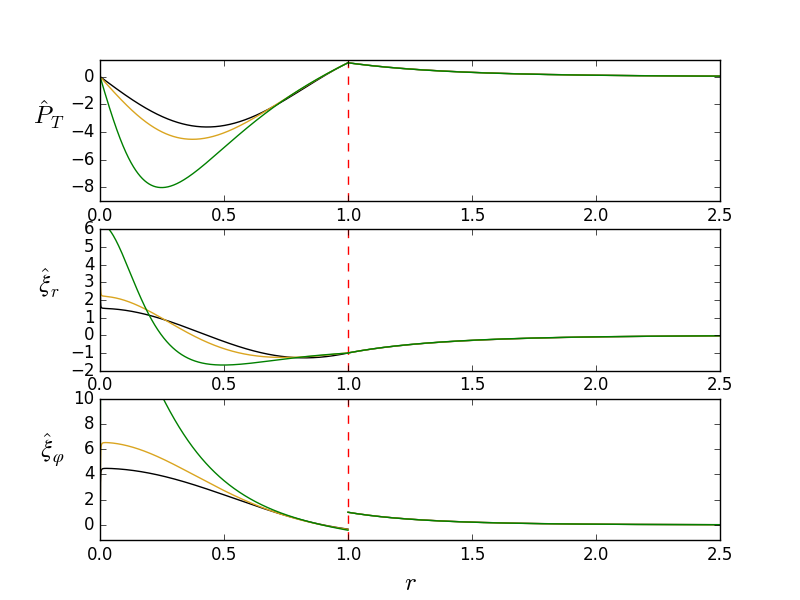}
        \caption{}
        \label{slow_body_kink_photospheric}
    \end{subfigure} 
    \caption{Resulting eigenfunctions for MHD wave modes in a photospheric cylinder with a non-uniform background plasma density given by the distributions shown in Figure (\ref{Gaussian_density_profiles_photospheric}) where the colour scheme is consistent. (a) Fast sausage surface mode for all cases with $ka=2.7$, (b) fast kink surface mode for all cases with $ka=1.3$, (c) slow sausage body mode with $ka=2.7$, (d) slow kink body mode for three cases all with $ka=3.2$. No azimuthal component is shown for the sausage mode as this wave mode does not produce an azimuthal perturbation. The case for $W=0.9$ not shown in (d) as this wave mode is cut off by the inhomogeneity. All plots are normalised to the external boundary value for each eigenfunction.}
    \label{photospheric_density_eigenfunctions}
   \end{figure*}
In this section a photospheric magnetic flux tube with a non-uniform background density profile is considered. Under photospheric conditions, the non-uniform background density is shown in Figure (\ref{Gaussian_density_profiles_photospheric}). Here, the centre of the flux tube is a local minimum for the internal density distribution, where the density increases towards the boundary at a rate which depends on the width of the Gaussian distribution. In all non-uniform cases the density at the boundary tends towards the external density. This also introduces bands on the dispersion diagrams for all characteristic speeds which depend on the plasma density. The internal sound, Alfv\'{e}n and tube speeds all now range from the value at the centre of the flux tube to the value at the boundary, denoted as $c_B$, $v_{AB}$ and $c_{TB}$. The resulting characteristic frequencies where the discrete wave modes are resonantly damped are shown in the appendix, in Figure~\ref{Photospheric_density_appendix}. As would be expected, the case for a large Gaussian width, Figure (\ref{Photospheric_density_dds_1e5}), corresponding to a uniform distribution produces the same dispersion diagram as shown for the uniform scenario in Figure (\ref{photospheric_uniform}). As the inhomogeneity is increased, the fast sausage and kink modes remains relatively unaffected, however are shifted to slightly slower phase speeds with increasing background non-uniformity. Figure (\ref{Photospheric_density_dds}) shows the behaviour of all wave modes as the inhomogeneity of background plasma density is increased. The slow body modes remain trapped between $c_{Ti}$ and $c_i$ although appear to also have slower phase speeds as the inhomogeneity is increased. For sufficient non-uniformity certain slow body modes can be cut off below $c_{Ti}$. Similar to Paper 1 these modes exist within a band shown by the green shaded region but this band does not represent a continuum therefore this is physically permittable. Furthermore, as the level of background density non-uniformity is increased, the slow surface modes in the photospheric case propagate with speeds similar to $c_{TB}$, which obviously changes with the inhomogeneity. At sufficiently large inhomogeneity, these modes disappear from the dispersion diagram. It is clear from both Figure (\ref{Coronal_density_dds}) and Figure (\ref{Photospheric_density_dds}) that a non uniform background density has the effect of increasing the propagating phase speeds of wave modes in a coronal cylinder and decreasing the phase speeds in a photospheric cylinder.

Turning attention now to the physical appearance of the eigenfunctions of the wave modes in a non-uniform photospheric magnetic cylinder. Figure (\ref{photospheric_density_eigenfunctions}) shows the spatial behaviour of fast surface and slow body modes for both the kink and sausage mode in a photospheric cylinder under all scenarios displayed in Figure (\ref{Gaussian_density_profiles_photospheric}). Similar to Paper 1, both the kink and sausage fast surface modes appear to be unaffected by the background inhomogeneity for $\hat{P}_T$ and $\hat{\xi}_r$. However Figure (\ref{photospheric_density_fast_kink}) shows that the azimuthal perturbation $\hat{\xi}_{\varphi}$ becomes more pronounced at the boundary as the background plasma inhomogeneity is increased. The slow body modes for both sausage and kink are shown in Figures (\ref{slow_body_sausage_photospheric}) and (\ref{slow_body_kink_photospheric}). Coinciding with the previously obtained results for a non-uniform magnetic slab, it was found that these modes were most affected by the background inhomogeneity, this is again true for a cylindrical waveguide. Both $\hat{P}_T$ and $\hat{\xi}_r$ are greatly affected for both sausage and kink modes and show the appearance of extra nodes and points of inflexion as the background inhomogeneity is increased. The azimuthal component $\hat{\xi}_{\varphi}$ also shows this behaviour for the slow kink mode. Finally, Figure (\ref{photospheric_density_eigenfunctions}) highlights the differences that a non-uniform background plasma density has on surface modes and body modes. Surface modes are defined as having a positive squared radial wavenumber and have maximum amplitude at the surface of the waveguide. Body modes have a negative squared radial wavenumber and as such exhibit oscillatory behaviour throughout the waveguide, possessing nodes inside the cylinder. Introducing a non-uniform background plasma density changes the spatial behaviour of surface modes at the boundary, most notably the azimuthal component, with the internal structure near the centre remaining locally constant. On the other hand a non-uniform background plasma density changes the local internal structure of the eigenfunctions for body modes, with the boundary values remaining unchanged no matter the scale of inhomogeneity of background equilibrium.

\subsection{Coronal conditions}\label{Coronal_plasma_density_section}

   \begin{figure}
   \centering
    \includegraphics[width=8.5cm]{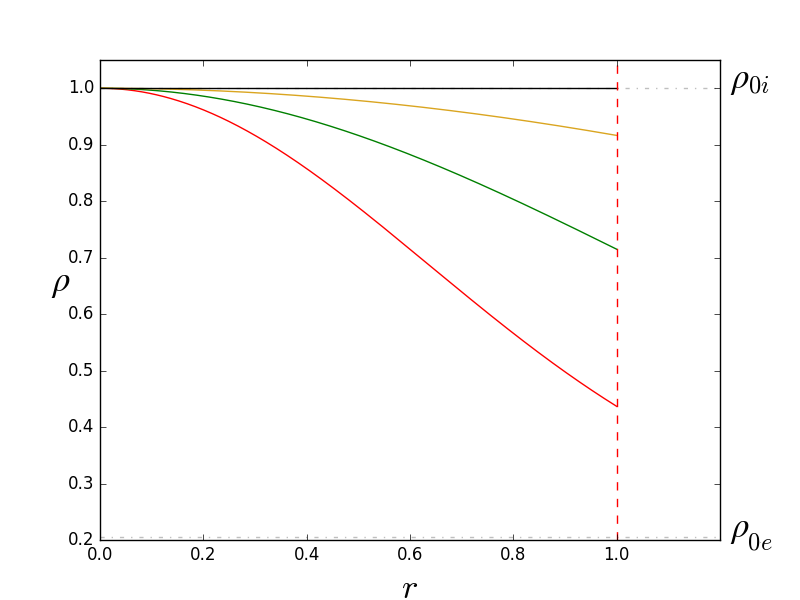}
    \caption{Gaussian background density profiles studied in this work for a cylinder under coronal conditions. $W = 10^5$ (black), $W = 3$ (yellow), $W = 1.5$ (green) and $W = 0.9$ (red).}
    \label{Gaussian_density_profiles}
   \end{figure}

   \begin{figure*}
   \centering
   \begin{subfigure}{.49\textwidth}
        \centering
        \includegraphics[width=9.cm]{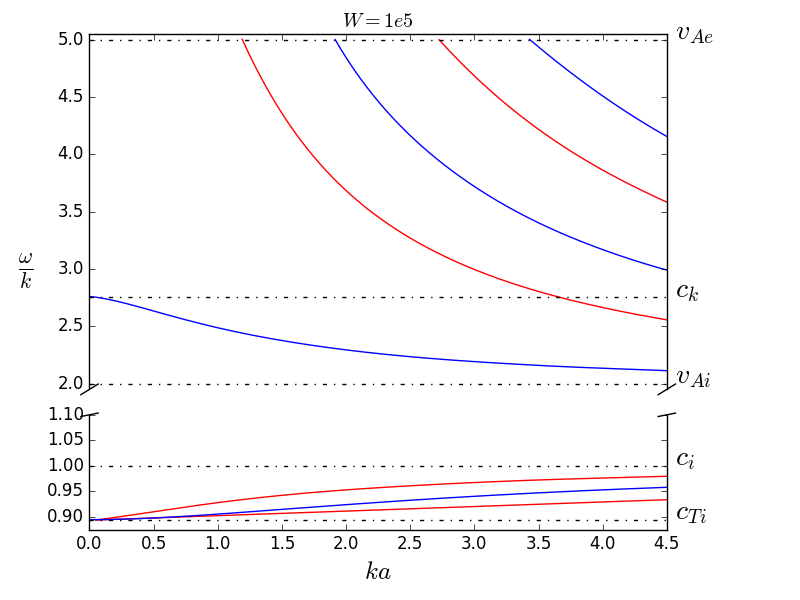}
        \caption{}
        \label{Coronal_density_dds_1e5}
    \end{subfigure}
   \begin{subfigure}{.49\textwidth}
        \centering
        \includegraphics[width=9.cm]{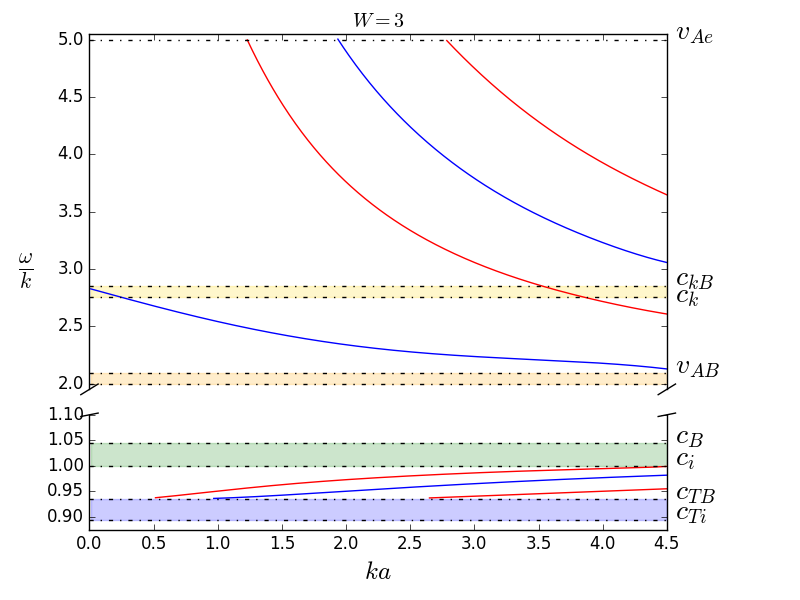}
        \caption{}
        \label{}
    \end{subfigure}   
   \begin{subfigure}{.49\textwidth}
        \centering
        \includegraphics[width=9.cm]{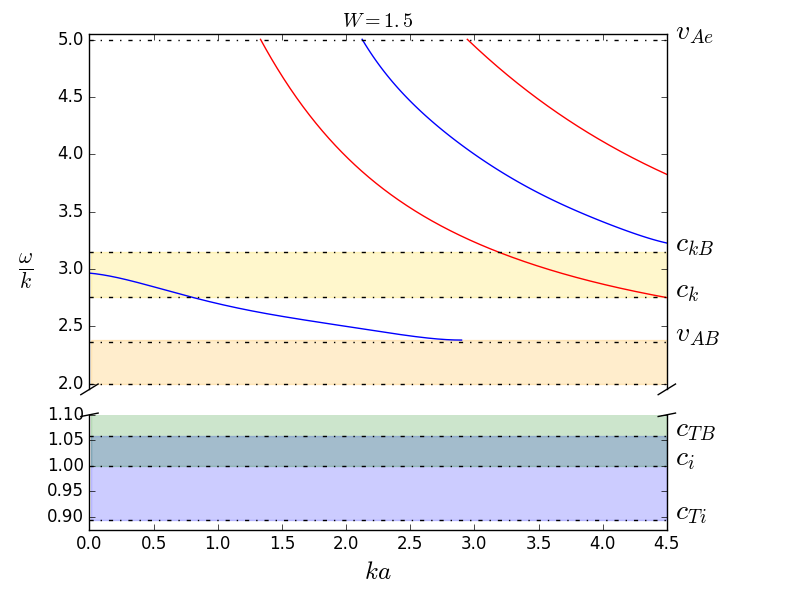}
        \caption{}
        \label{}
    \end{subfigure} 
   \begin{subfigure}{.49\textwidth}
        \centering
        \includegraphics[width=9.cm]{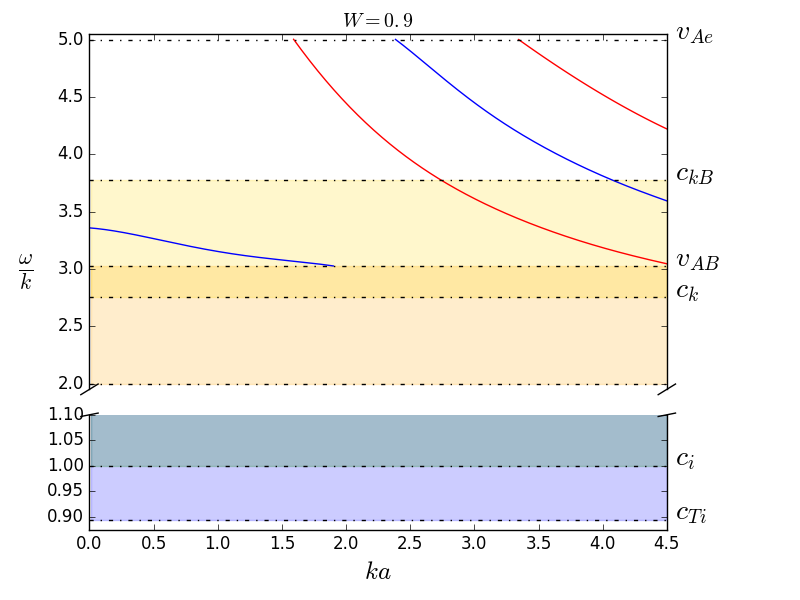}
        \caption{}
        \label{}
    \end{subfigure} 
    \caption{Dispersion diagrams for magnetoacoustic waves in a coronal cylinder with a background plasma density in the form of Gaussian profiles shown in Figure (\ref{Gaussian_density_profiles}). (a) $W=10^5$ corresponding to a uniform flow, (b) $W=3$, (c) $W=1.5$ and (d) $W=0.9$. Red curves denote sausage mode, blue curves show kink mode. Shaded regions represent the non-uniform bands due to the equilibrium inhomogeneity. The slow continuum (blue shaded region), inhomogeneous sound speed band (green shaded region), Alfv\'{e}n continuum (pink shaded region) and inhomogeneous kink speed band (orange shaded region) are all shown.}
    \label{Coronal_density_dds}
   \end{figure*}

   \begin{figure*}
   \centering
   \begin{subfigure}{.49\textwidth}
        \centering
        \includegraphics[width=9.cm]{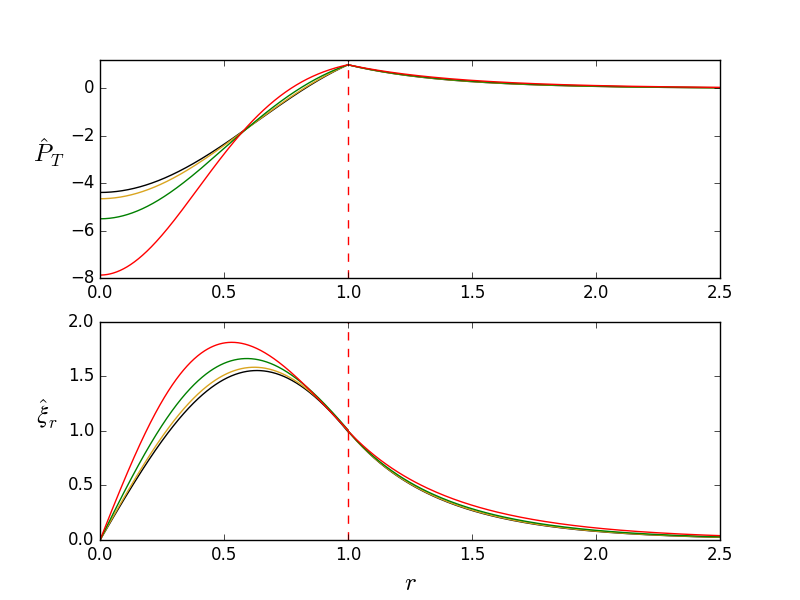}
        \caption{}
        \label{Coronal_density_eigenfunc_fast_sausage}
    \end{subfigure}
   \begin{subfigure}{.49\textwidth}
        \centering
        \includegraphics[width=9.cm]{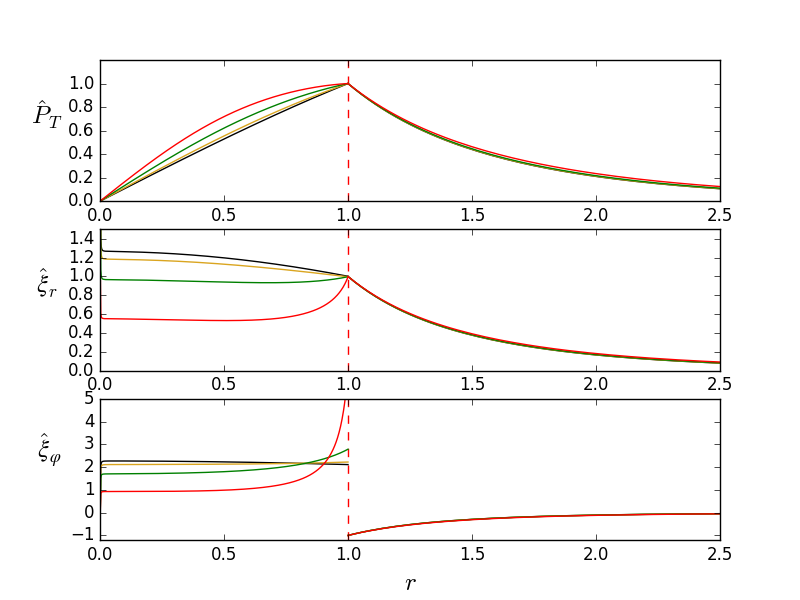}
        \caption{}
        \label{Coronal_density_eigenfunc_fast_kink}
    \end{subfigure}   

    \caption{Resulting eigenfunctions for MHD wave modes in a coronal cylinder with a non-uniform background plasma density given by the distributions shown in Figure (\ref{Gaussian_density_profiles}) where the colour scheme is consistent. (a) Fast sausage body mode for all cases with $ka=2.75$, (b) fundamental kink body mode for all cases with $ka=1.5$. No azimuthal component is shown for the sausage mode as this wave mode does not produce an azimuthal perturbation in this case. All plots are normalised to the external boundary value for each eigenfunction.}
    \label{Coronal_density_eigenfunctions}
   \end{figure*}

   \begin{figure*}
   \centering
   \begin{subfigure}{.49\textwidth}
        \centering
        \includegraphics[width=8.6cm]{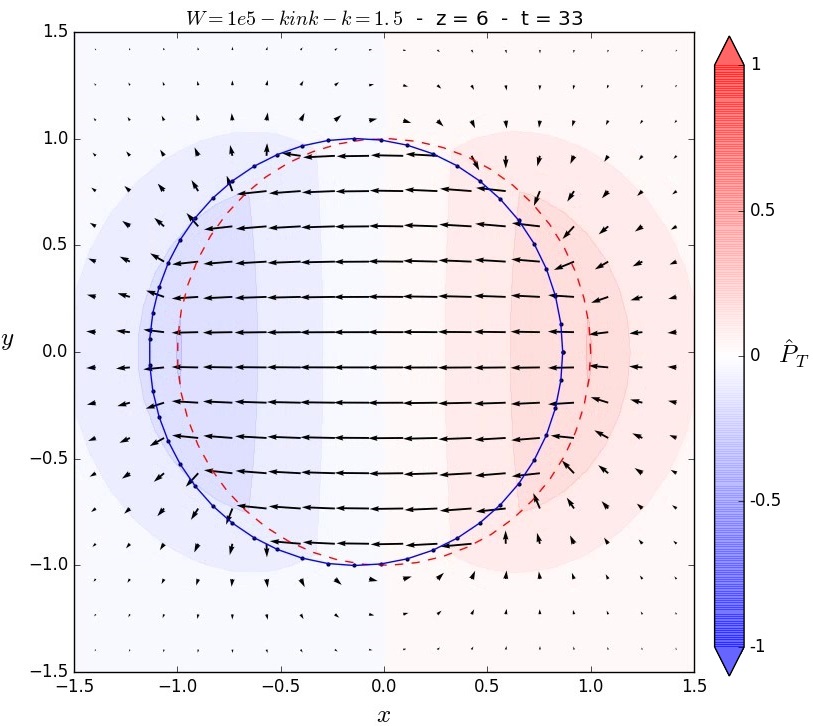}
        \caption{}
        \label{Uniform_Density_velocity_field}
    \end{subfigure}
   \begin{subfigure}{.49\textwidth}
        \centering
        \includegraphics[width=8.6cm]{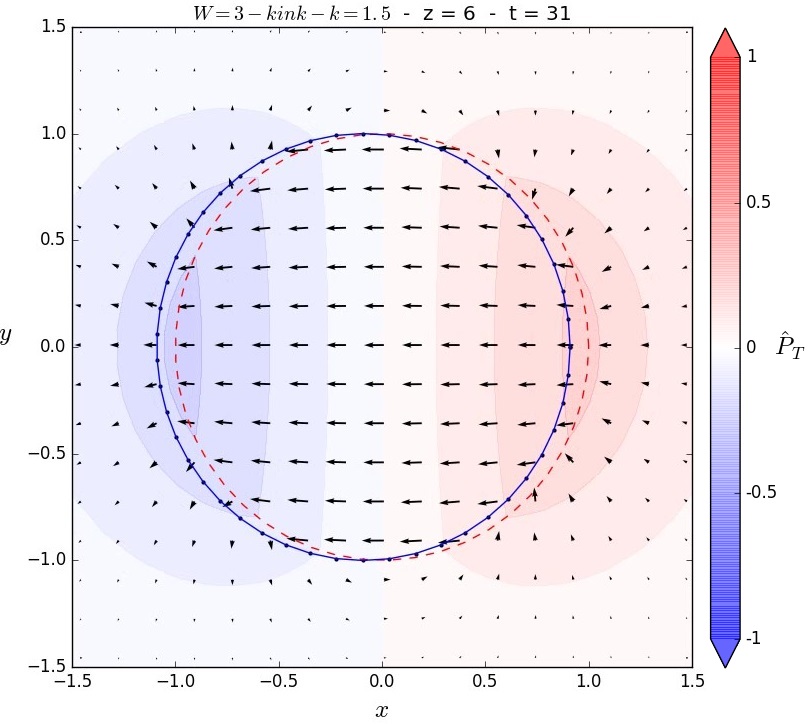} 
        \caption{}
    \end{subfigure} 
       \begin{subfigure}{.49\textwidth}
        \centering
        \includegraphics[width=8.6cm]{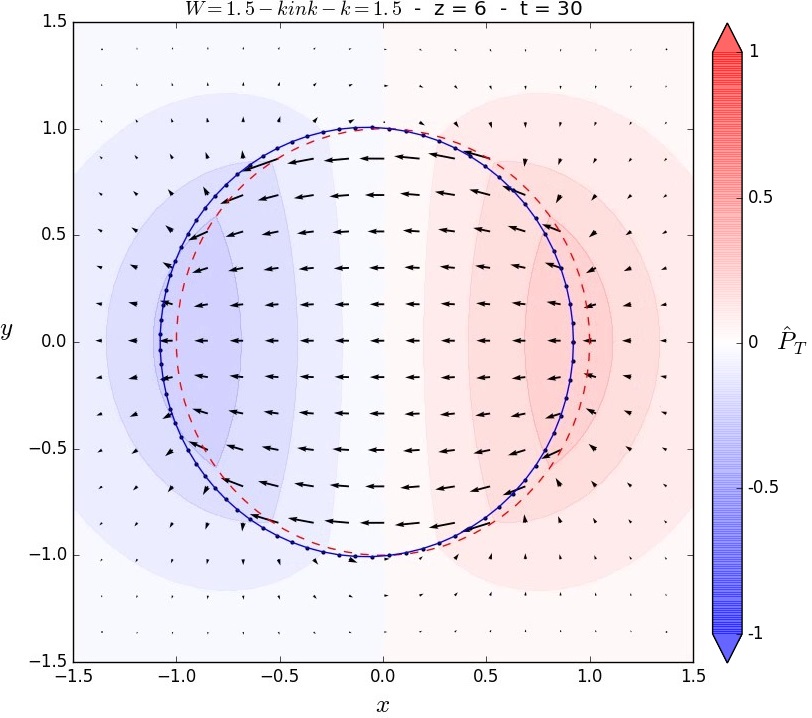}
        \caption{}
    \end{subfigure}
       \begin{subfigure}{.49\textwidth}
        \centering
        \includegraphics[width=8.6cm]{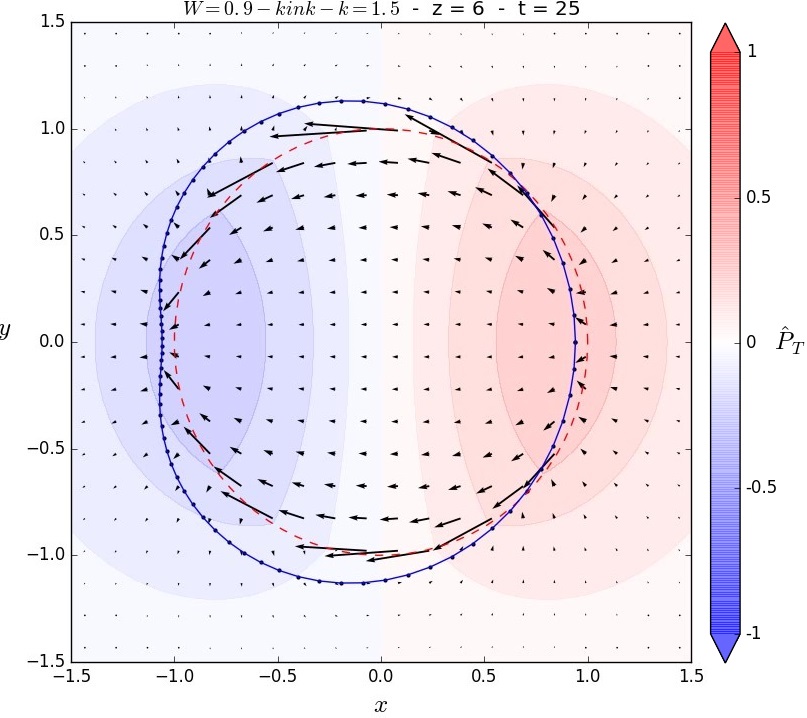}
        \caption{}
        \label{W09_Density_velocity_field}
    \end{subfigure}

    \caption{Snapshots of the velocity field in time at the moment of maximum displacement for the different Gaussian profiles modelling plasma density for the fast fundamental kink mode. (a) $W = 10^5$, (b) $W = 3$, (c) $W = 1.5$ and (d) $W = 0.9$. The eigenfunctions shown in Figure (\ref{Coronal_density_eigenfunc_fast_kink}) are converted into Cartesian components and shown in a Cartesian grid. The same value of $ka = 1.5$ is chosen in all plots. The colour contour shows the normalised total pressure perturbation where blue is negative and red is positive. The solid blue line outlines the shape of the perturbed boundary. Seen in Figure (\ref{W09_Density_velocity_field}) is the linear regime of a similar case study conducted by \citet{ant2014} (see their Figure 1) in which the authors investigate the non-linear Kelvin-Helmholtz instability modelling a non-uniform transition layer in a magnetic cylinder.}
    \label{Gaussian_density_velocity_field}
   \end{figure*}

   \begin{figure*}
   \centering
   \begin{subfigure}{.49\textwidth}
        \centering
        \includegraphics[width=10.5cm]{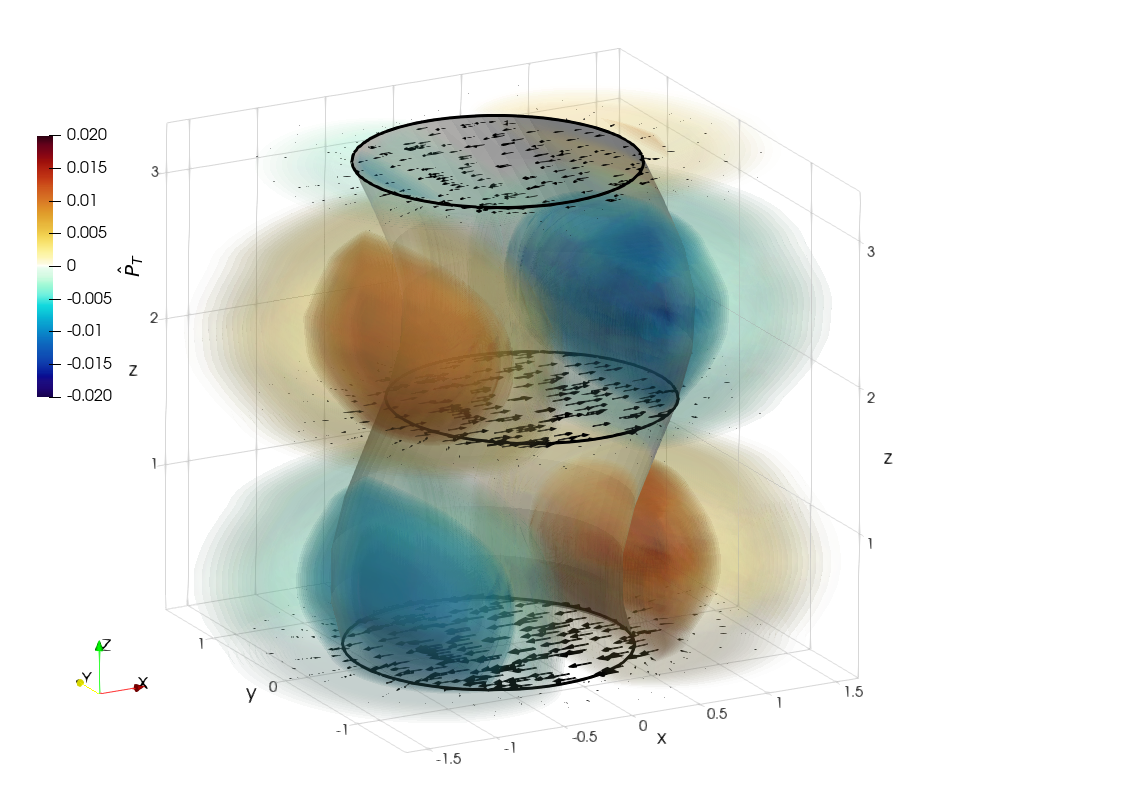}
        \caption{}
        \label{3D_Gaussian_density_Suzana_uniformdensity}
    \end{subfigure}
   \begin{subfigure}{.49\textwidth}
        \centering
        \includegraphics[width=10.5cm]{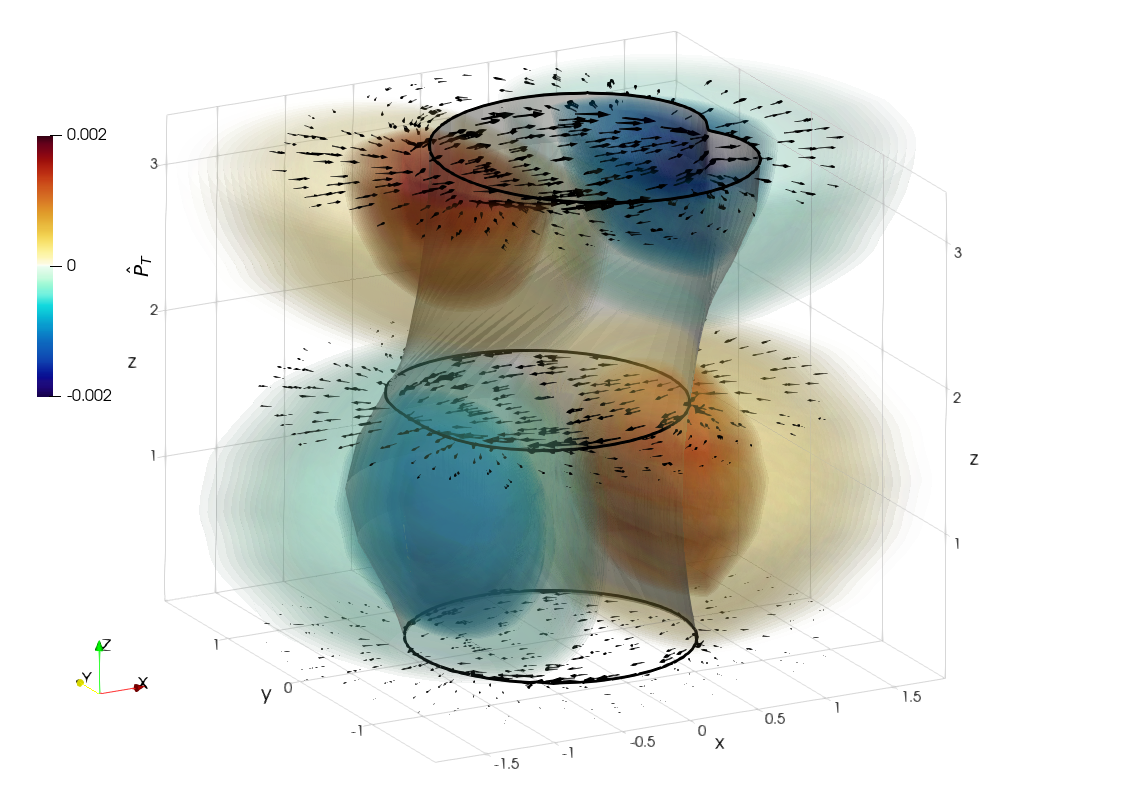}
        \caption{}
        \label{3D_Gaussian_density_Suzana_W09}
    \end{subfigure}   

    \caption{3D visualisation of $\hat{P}_T$ and the perturbed velocity vector field in the presence of a uniform and non-uniform background plasma density for the fundamental kink mode with eigenfunctions shown in Figure (\ref{Coronal_density_eigenfunc_fast_kink}). These correspond to the 2D velocity field vectors shown in Figure (\ref{Uniform_Density_velocity_field}) and Figure (\ref{W09_Density_velocity_field}) respectively. (a) Case for uniform plasma density (b) case with Gaussian density with $W=0.9$. Movies of these 3D visualisations can be found online at the PDG visualisations web-page.}
    \label{3D_Gaussian_density_Suzana}
   \end{figure*}

    

Unlike the similar case studied in Paper 1 for a magnetic slab, the characteristic speeds chosen to represent coronal conditions are slightly changed in the case of a cylindrical magnetic flux tube. This changes the characterstics of the dispersion diagram, namely that the slow body modes experience a cut off at certain values of inhomogeneity, which wasn't present in Paper 1 due to the smaller speed of $v_{Ae}$ used in the analysis. The background density profiles investigated in this paper are the same as Paper 1 and displayed in Figure (\ref{Gaussian_density_profiles}) for increasing non-uniformity where the internal density gradually tends towards the external density at the boundary. The resulting characteristic frequencies where the discrete wave modes are resonantly damped in the non-uniform coronal cylinder cases investigated in this section are shown in the appendix, in Figure~\ref{coronal_density_appendix}.
Figure (\ref{Coronal_density_dds}) shows the behaviour of sausage and kink modes in a coronal cylinder with a background spatial density profile as a Gaussian distribution. The case of a large width if the Gaussian distribution is given by $W=10^5$ and the resulting dispersion diagram is shown in Figure (\ref{Coronal_density_dds_1e5}). As expected, this case produces the exact result as the uniform investigation by \citet{edrob1983} and shown in this work in Figure (\ref{coronal_uniform}). As the inhomogeneity of the background plasma density is increased, the density value at the boundary becomes smaller (tends towards the external value however is still discontinuous across $r=a$). As a result, the variables which depend on density such as $c_i$, $v_{Ai}$ and $c_{Ti}$ become a continuous band across multiple possible phase speeds, these are shown by the shaded regions in Figure (\ref{Coronal_density_dds}). It is well known that the Alfv\'{e}n and cusp continuum are regions in which dissipative processes are possible, such as phase mixing and resonant absorption, due to local resonances occurring within these bands. The wave frequency becomes complex here and as such the real part of the phase speed is cut off on the dispersion diagram by $v_{AB}$ and $c_{TB}$. In a cylinder with a large enough inhomogeneity of plasma density, the slow body modes disappear from the dispersion diagram as no real band exists in which they can propagate, similar to the results discussed in Paper 1.

In a non-uniform plasma, it is well known that a smooth inhomogeneity such that the density varies linearly from one value ($\rho_1$) to another ($\rho_2$), that quasi-modes are introduced \citep{Sed1971, Tirry1996, Priest2014}. The real part of the quasi mode phase speed is defined as $\sqrt{(\rho_1 v_{A1}^2 + \rho_2 v_{A2}^2)/(\rho_1 + \rho_2)}$ which is simply the kink speed between the minimum and maximum value. Replacing the smooth linear non-uniform density by an external $\rho_{0e}$ and internal $\rho_{0i}$ value yields the well known kink speed for a uniform cylinder. Seen in Figure (\ref{Coronal_density_dds}), the fundamental kink branch does not tend to either of these values in the thin-tube limit. Instead, it tends towards a value in between which is due to the fact that the inhomogeneity is not smooth from outside to inside the cylinder, in all cases the density is discontinuous across $r=a$.

Figure (\ref{Coronal_density_eigenfunctions}) displays the spatial eigenfunctions for both the fast sausage mode and fundamental kink mode in a coronal cylinder with a non-uniform background density. Paper 1 concluded that fast modes are unaffected by the background inhomogeneity in a magnetic slab, however their behaviour may be different in a magnetic cylinder. It can be seen clearly in Figure (\ref{Coronal_density_eigenfunctions}) that as the plasma density inhomogeneity is increased, both the fast sausage mode and fundamental kink mode display different spatial characteristics for both $\hat{P}_T$ and $\hat{\xi}_r$ and also $\hat{\xi}_{\varphi}$ for the kink mode. The fundamental kink mode was not present in the magnetic slab analysis and this is the mode which is routinely observed in the thin-tube limit in the solar atmosphere when a cylindrical analytical model is considered. The difference in spatial eigenfunction behaviour is also more pronounced for $ka < 1$ for the fundamental kink mode. As the fast sausage mode experiences a cut off around $ka=1.5$, it is unlikely that these eigenfunctions would be observed for this mode in non-uniform coronal structures.

A snapshot in time of the resulting velocity field at maximum displacement for the fundamental kink mode is shown in Figure (\ref{Gaussian_density_velocity_field}) for all cases of equilibrium Gaussian density. The velocity field corresponds to the eigenfunctions shown in Figure (\ref{Coronal_density_eigenfunc_fast_kink}) converted into Cartesian components to be visualised in a uniform Cartesian grid. The case for uniform density retrieves the theoretical kink mode displacement in that the velocity field is uniform inside the cylinder and has a dipole configuration in the external region as seen in Figure (\ref{Uniform_Density_velocity_field}). As the equilibrium inhomogeneity is increased, the resulting velocity field inside the cylinder becomes curved and the azimuthal component dominates. For the case of maximum inhomogeneity given by $W=0.9$ shown in Figure (\ref{W09_Density_velocity_field}), this increasing azimuthal component results in the boundary of the cylinder becoming distorted. This result can be understood by looking at the azimuthal component of the eigenfunction in Figure (\ref{Coronal_density_eigenfunc_fast_kink}) where the magnitude of discontinuity at the boundary increases with increasing inhomogeneity. The nature of the azimuthal displacement component for the kink mode can be understood by examining Equation (\ref{azimuthal_comp}) when the location $r=1$ is crossed. At this position, $\omega^2 -k^2v_A^2$ changes sign discontinuously whereas the total pressure perturbation $\hat{P}_T$ remains continuous. Furthermore, in the case for maximum inhomogeneity, the frequency $\omega$ approaches the local resonant Alfv\'{e}n frequency $k^2 v_A^2$ which results in the large amplitude for $\hat{\xi}_{\varphi}$. The increased discontinuity in displacement creates counter-streaming flows that can generate the Kelvin-Helmholtz instability.  These results can be compared to the linear stage of \citet{ant2014} (see their Figure 1) in which similar behaviour of the boundary is seen but in the case of a thinner boundary layer with a non-uniform density profile. This behaviour has also been detected in previous numerical studies investigating straight cylinders with a non-uniform density layer in the radial direction \citep{Terradas2008} including non-ideal MHD \citep{Howson2017} and also an analytic study with a velocity shear in the azimuthal component across the boundary \citep{sol2010}. Figure (\ref{3D_Gaussian_density_Suzana}) shows a 3D representation of the 2D velocity fields seen in Figure (\ref{Gaussian_density_velocity_field}) for a propagating wave in vertical coordinate $z$. It can be clearly seen in Figure (\ref{3D_Gaussian_density_Suzana_W09}) that at maximum displacement for the kink mode, the boundary becomes distorted due to the non-uniform equilibrium plasma density. This perturbation of the boundary propagates with the wave vertically through the magnetic flux tube, unlike the uniform scenario shown in Figure (\ref{3D_Gaussian_density_Suzana_uniformdensity}) which maintains the structure of the tube. It was suggested that the fundamental kink mode in a non-uniform plasma should actually be called a surface Alfv\'{e}n wave due to the mixed properties and increased vorticity, which is not a property associated with magnetoacoustic waves \cite{goo2012}. The results presented in this section further strengthen this debate as the kink mode does not display traditional properties when the equilibrium plasma is non-uniform.

\section{Inhomogeneous field aligned flow}\label{gaussian_flow}
   \begin{figure}
   \centering
    \includegraphics[width=7.5cm]{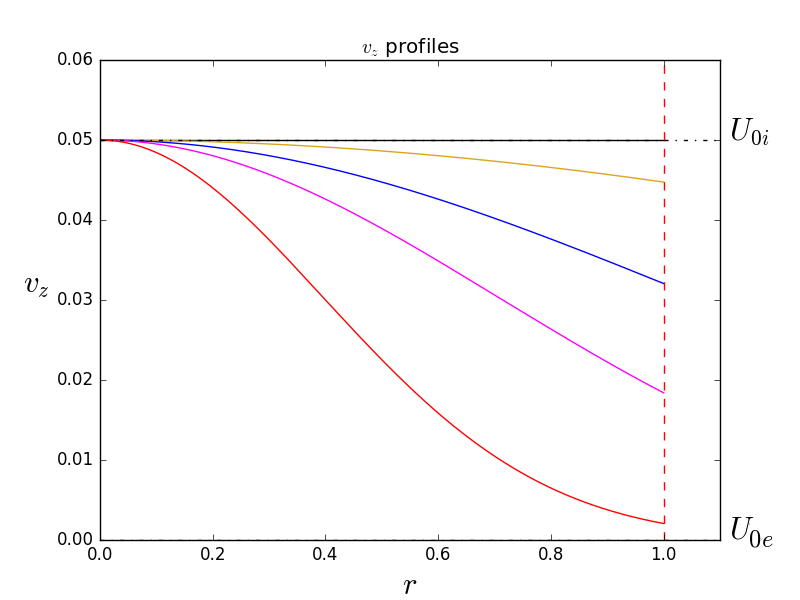}
    \caption{Gaussian flow profiles inside an otherwise uniform coronal magnetic cylinder. $W = 10^5$ (black), $W = 3$ (yellow), $W = 1.5$ (blue), $W = 1$ (magenta) and $W = 0.6$ (red).}
    \label{Gaussian_flow_profiles}
   \end{figure}

   \begin{figure*}
   \centering
   \begin{subfigure}{.49\textwidth}
        \centering
        \includegraphics[width=9.cm]{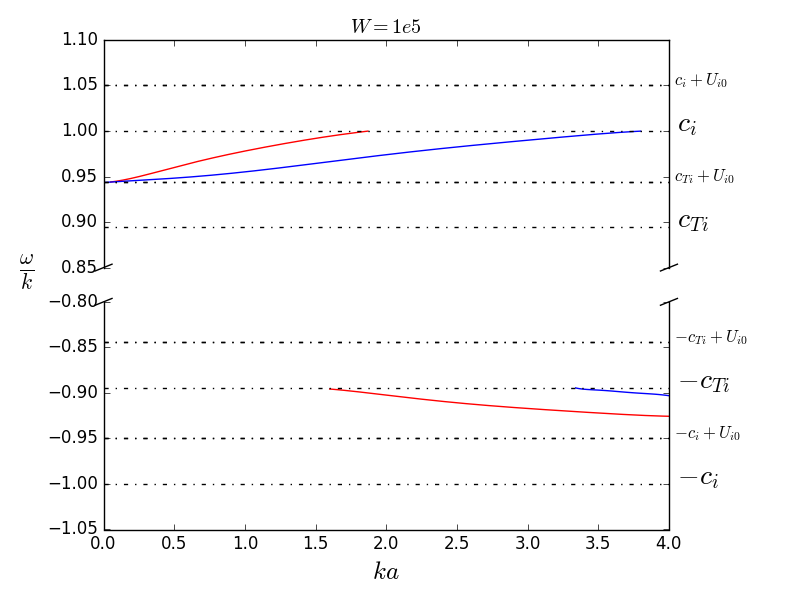}
        \caption{}
        \label{}
    \end{subfigure}
   \begin{subfigure}{.49\textwidth}
        \centering
        \includegraphics[width=9.cm]{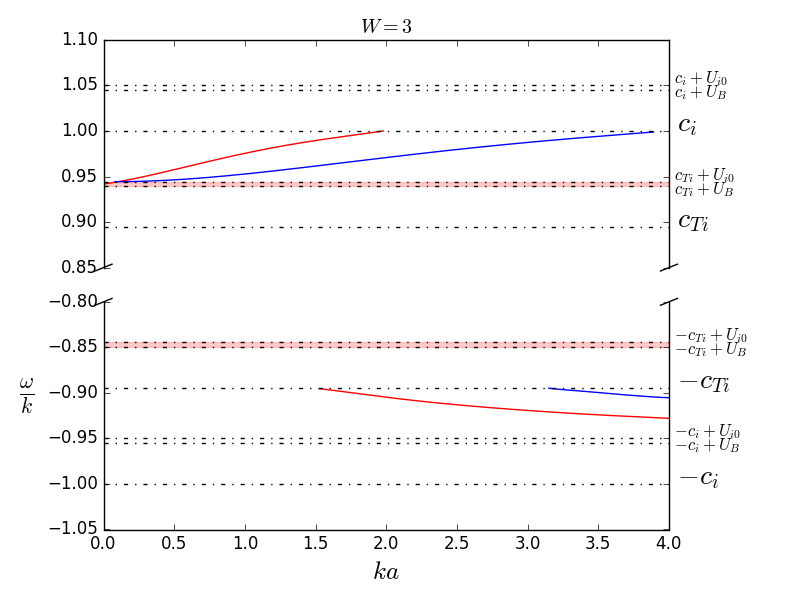}
        \caption{}
        \label{}
    \end{subfigure}   
   \begin{subfigure}{.49\textwidth}
        \centering
        \includegraphics[width=9.cm]{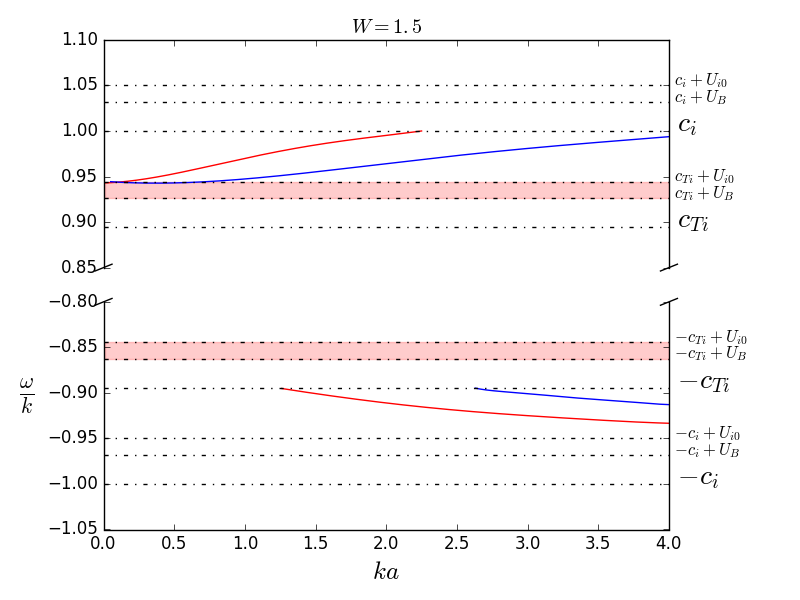}
        \caption{}
        \label{}
    \end{subfigure} 
   \begin{subfigure}{.49\textwidth}
        \centering
        \includegraphics[width=9.cm]{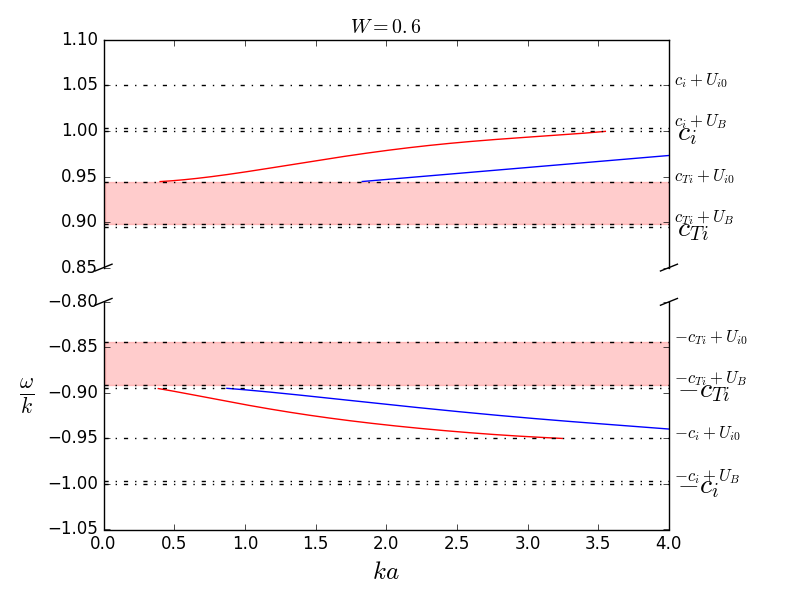}
        \caption{}
        \label{}
    \end{subfigure} 
    \caption{Zoom in on dispersion diagrams for forward and backward propagating slow body modes in a coronal cylinder with a background plasma flow in the form of Gaussian profiles shown in Figure (\ref{Gaussian_flow_profiles}). (a) $W=10^5$ corresponding to a uniform flow, (b) $W=3$, (c) $W=1.5$ and (d) $W=0.6$. Red curves denote sausage mode, blue curves show kink mode. The red shaded bands indicate the flow continuum where the modes become resonantly damped.}
    \label{Gaussian_flow_dds}
   \end{figure*}

   \begin{figure*}
   \centering
   \begin{subfigure}{.49\textwidth}
        \centering
        \includegraphics[width=9.cm]{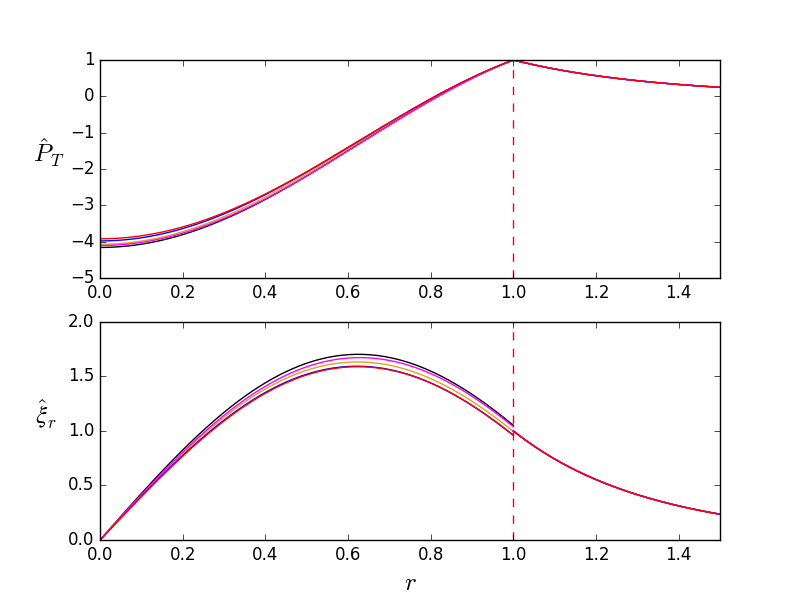}
        \caption{}
        \label{}
    \end{subfigure}
   \begin{subfigure}{.49\textwidth}
        \centering
        \includegraphics[width=9.cm]{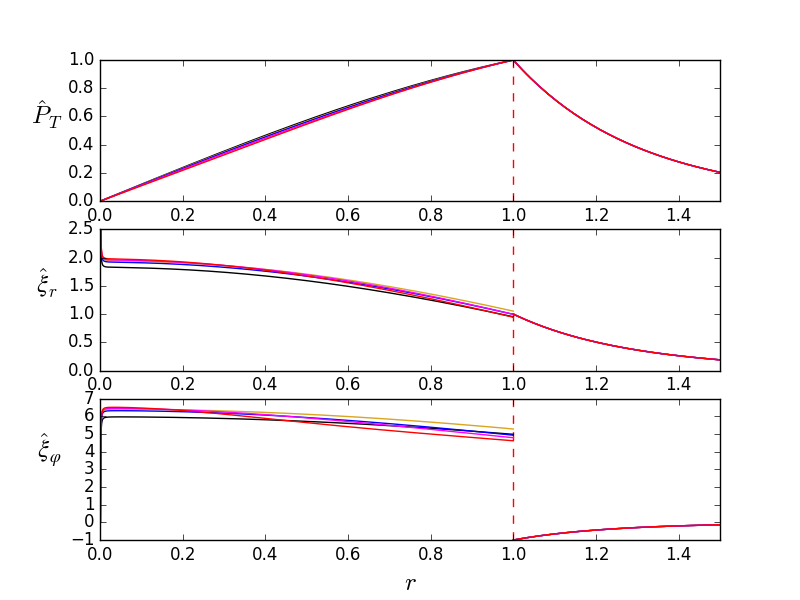}
        \caption{}
        \label{}
    \end{subfigure}   
   \begin{subfigure}{.49\textwidth}
        \centering
        \includegraphics[width=9.cm]{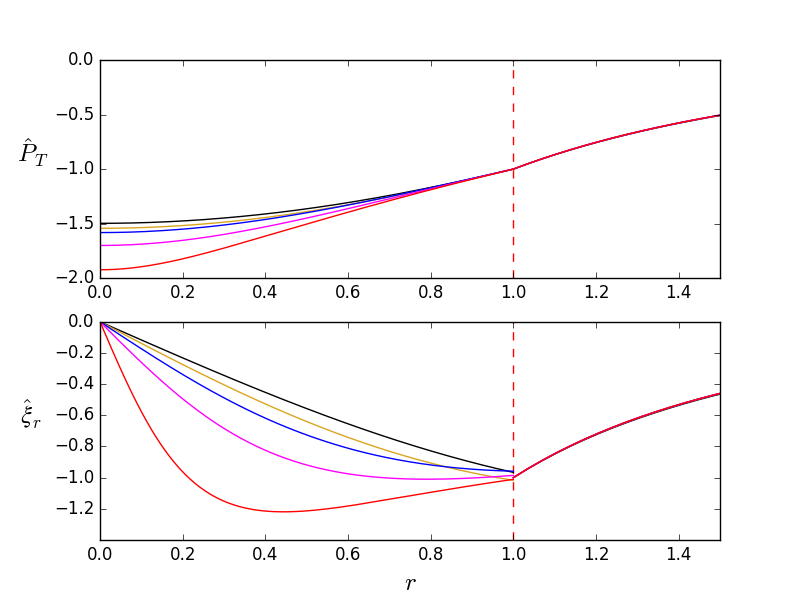}
        \caption{}
        \label{}
    \end{subfigure} 
   \begin{subfigure}{.49\textwidth}
        \centering
        \includegraphics[width=9.cm]{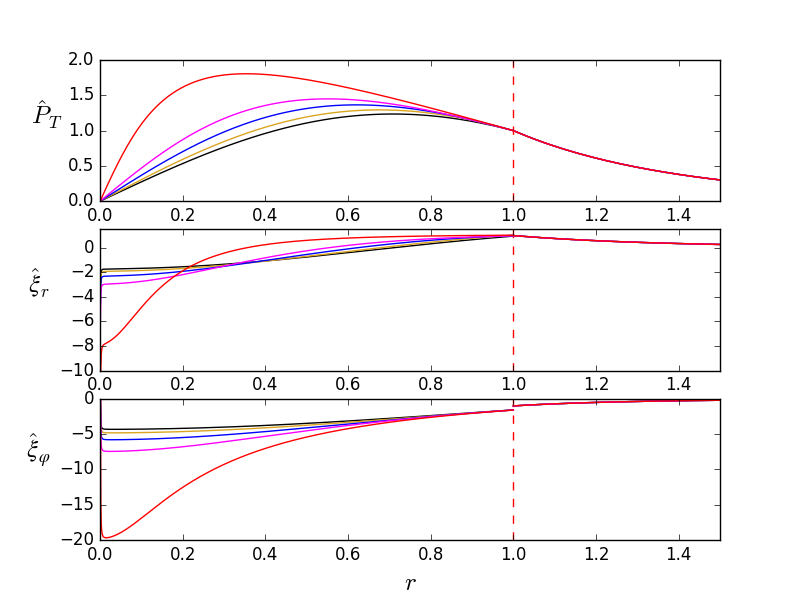}
        \caption{}
        \label{gaussian_flow_eigenfunction_slowkink}
    \end{subfigure} 
    \caption{Eigenfunctions for a coronal cylinder with a background Gaussian flow as shown in Figure (\ref{Gaussian_flow_profiles}) where the colour scheme is consistent with the equilibrium profiles. (a) Fast sausage body mode with $ka=3$, (b) fast kink body mode with $ka=3$, (c) slow sausage body mode with $ka=1$, (d) slow kink body mode with $ka=2$.}
    \label{Gaussian_flow_eigenfunctions}
   \end{figure*}

   \begin{figure*}
   \centering
    \includegraphics[width=15.65cm]{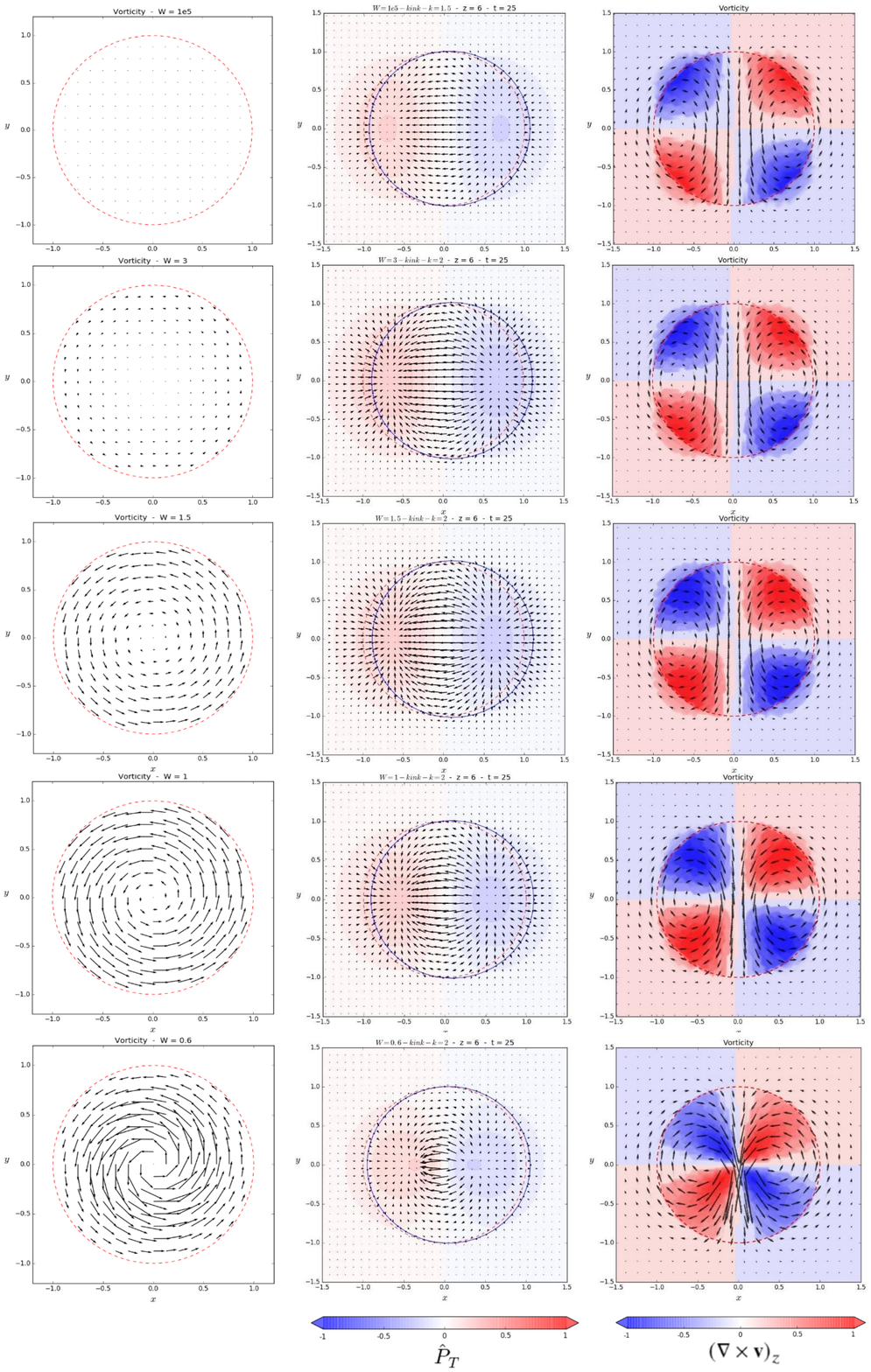}
    \caption{Background $x-y$ vorticity field (left), perturbed $x-y$ velocity field (middle) and background plus perturbed $x-y$ vorticity field (right) plots for all cases of Gaussian plasma flow shown in Figure (\ref{Gaussian_flow_profiles}). These snapshots all correspond to the slow body kink mode with eigenfunctions shown in Figure (\ref{gaussian_flow_eigenfunction_slowkink}). Top row corresponds to $W = 10^5$ with inhomogeneity increasing down the plot through $W = 3$ (second row), $W = 1.5$ (third row), $W = 1$ (fourth row) to bottom row where $W = 0.6$. The colour contour in centre plots shows the total pressure perturbation whereas the colour contours in the right column plots denote the vorticity component perpendicular to the $x-y$ plane.}
    \label{Gaussian_flow_vorticity_comparisons}
   \end{figure*}

   \begin{figure*}
   \centering
   \begin{subfigure}{.49\textwidth}
        \centering
        \includegraphics[width=12.cm]{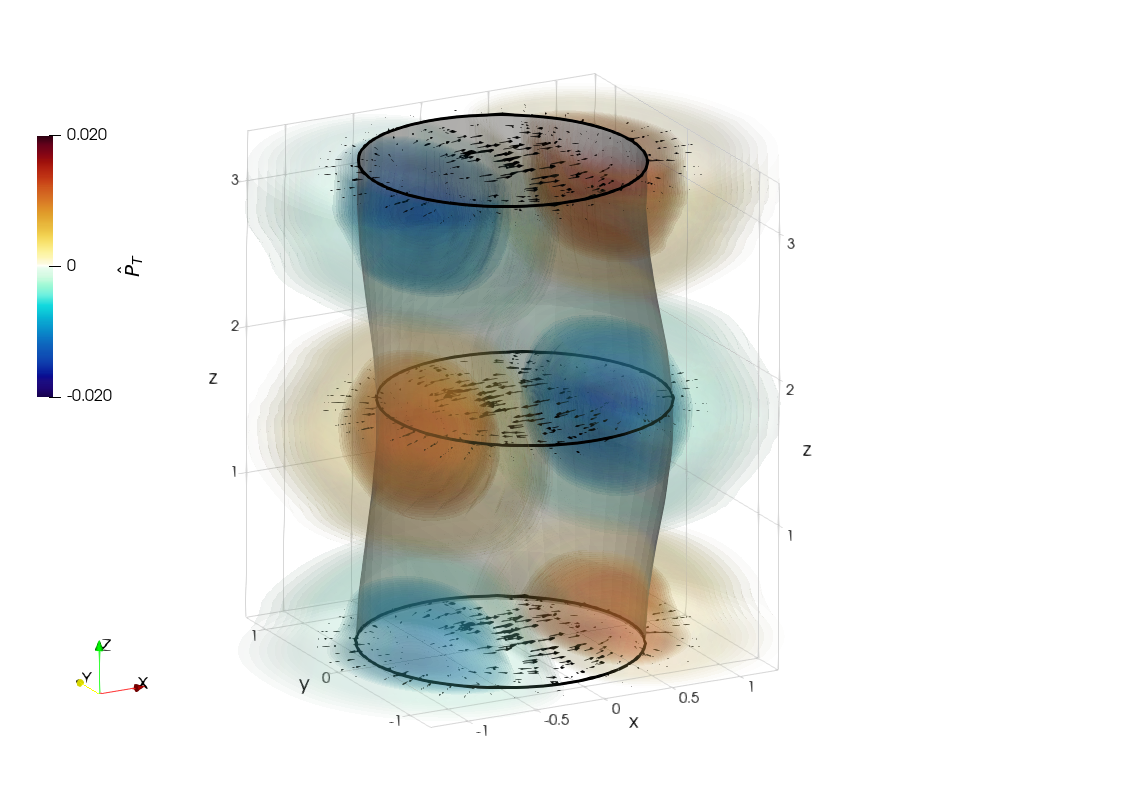}
        \caption{}
        \label{3D_Gaussian_flow_Suzana_uniformflow}
    \end{subfigure}
   \begin{subfigure}{.49\textwidth}
        \centering
        \includegraphics[width=11.cm]{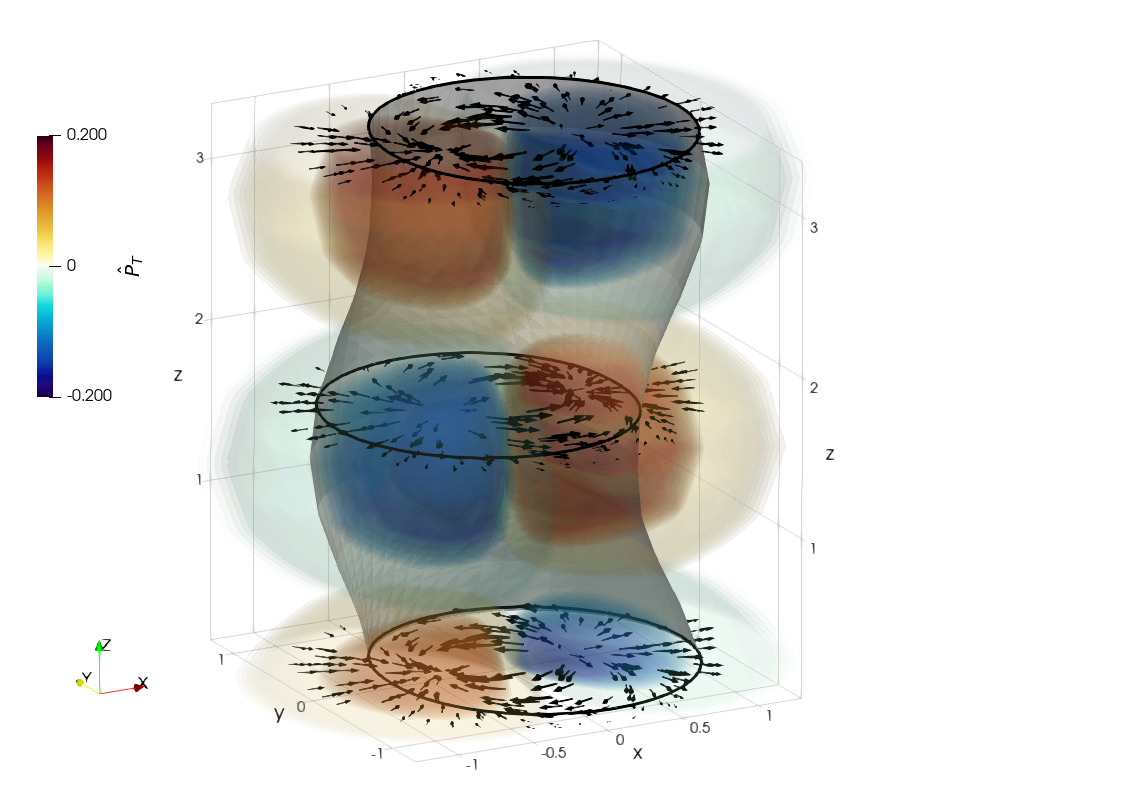}
        \caption{}
        \label{3D_Gaussian_flow_Suzana_W06}
    \end{subfigure}   

    \caption{
    3D visualisation of $\hat{P}_T$ and the perturbed velocity vector field in the presence of a uniform and non-uniform background plasma flow for the slow body kink mode with eigenfunctions shown in Figure (\ref{gaussian_flow_eigenfunction_slowkink}). These correspond to the 2D velocity field vectors shown in the middle panel of Figure (\ref{Gaussian_flow_vorticity_comparisons}). (a) Case for uniform plasma flow ($W=10^5$) (b) case with Gaussian flow with $W=0.6$. Movies of these 3D visualisations can be found online on the PDG visualisations web-page.}
    \label{3D_Gaussian_flow_Suzana}
   \end{figure*}

In this section, a magnetic cylinder of uniform plasma is modelled with a vertical straight magnetic field with a field aligned and radially non-uniform internal background plasma flow embedded in a coronal environment. A cartoon of this equilibrium configuration is shown by panel (c) in Figure (\ref{cylinder_cartoon}). A similar case study was investigated in Paper 1 in a Cartesian geometry. The flow magnitude is chosen to be small in comparison with the internal sound speed, $U_{0i} = 0.05c_i$. This allows a clearer investigation into the physical effects of the spatial flow to be undertaken as a large flow magnitude will shift certain wave modes into the leaky regime, which is not analysed in this work. Adopting a small magnitude of the background plasma flow speed also avoids the possibility of the onset of flow related instabilities such as Kelvin-Helmholtz. Similar to the previous investigation of a non-uniform plasma density, the background plasma flow in this section is also modelled as a series of Gaussian profiles. These profiles use the expression:
\begin{displaymath}
    U_{0i}(r) = U_{0e} + \left(U_{0i} - U_{0e}\right)\text{exp}\left(-\frac{(r-r_0)^2}{W^2}\right),
\end{displaymath}
where $U_{0e}$ is the flow outside the waveguide, assumed to be $0$.

These inhomogeneous flow profiles investigated are shown in Figure (\ref{Gaussian_flow_profiles}). The case of a large width (i.e $W = 10^5$) corresponds to a uniform steady flow which is well known to create an observed Doppler shift to the waves in the direction of flow \citep{nak1995}. Due to the small amplitude of flow chosen, it is found that there is very little effect on the fast modes in the dispersion diagram, there is a more observable affect on the slow body modes. A zoom in region of the resulting dispersion diagrams are shown in Figure (\ref{Gaussian_flow_dds}) for the forward and backward propagating slow body modes. It can be seen that as expected, these waves are shifted with respect to the maximum flow speed. As the non-uniformity of flow is increased, the branches of the forward and backward slow body modes are shifted with a clear asymmetry between the forward and backward propagating modes. With increasing spatial non-uniformity of the background plasma flow, the permittable backward propagating kink and sausage slow body modes propagate in the thin tube limit, whereas the opposite is true for forward propagating slow body modes. This effect is due to the flow speed at the boundary $U_B$ which depends on the initial non-uniformity of the background plasma flow, therefore a non-uniform background plasma flow may further shift some wave modes into possible propagation windows. The red shaded bands in Figure (\ref{Gaussian_flow_dds}) indicate the Doppler shifted continua given by Equation (\ref{cusp_flow_continuum}) where the wave modes become resonantly damped. It can be seen that as the inhomogeneity of the background flow is increased, these continuum bands become wider, providing a larger frequency domain for resonant processes to occur.

The resulting eigenfunctions of $\hat{P}_T$, $\hat{\xi}_r$ and $\hat{\xi}_{\varphi}$ are shown in Figure (\ref{Gaussian_flow_eigenfunctions}). It can be seen that fast modes are not heavily affected by the inhomogeneity of the equilibrium background plasma flow. This is mainly down to the before-mentioned amplitude of the non-uniform flow. The background plasma flow has the affect of Doppler shifting the waves which is much less clear for the dispersive fast waves. Unlike in Section \ref{Coronal_plasma_density_section}, where the amplitude of plasma density non-uniformity was large, here the equilibrium plasma is uniform. Slow body modes, however, feel the non-uniformity much more greatly. The local perturbation amplitude for all eigenfunctions is increased with the non-uniformity of the background flow and extra nodes and points of inflexion become visible. Furthermore the maximum local azimuthal perturbation $\hat{\xi}_{\varphi}$ is increased with a more inhomogeneous background plasma flow.

Another quantity which can be investigated is vorticity defined in this work as the curl of the velocity field, $\bf{\nabla} \times \bf{v}$. Vorticity plays an important role in the dynamics of the solar atmosphere. Granular motions in the photosphere produce a ubiquitous number of observed vortices in intergranular lanes. These vortices can have the effect of twisting the magnetic field lines which are rooted into the photosphere and exciting torsional Alfv\'{e}n waves \citep{fed2011, fedverth2011, Vig2012, Moll2012, Shel2013, Silva2020}. The background vorticity field, perturbed velocity field and the background plus perturbed vorticity field are plotted in Figure (\ref{Gaussian_flow_vorticity_comparisons}) for the slow body kink mode with eigenfunctions shown in Figure (\ref{gaussian_flow_eigenfunction_slowkink}). The left hand side column shows the background vorticity due to the equilibrium background plasma flow. Obviously, with a uniform bulk flow, there is no inhomogeneity and as a result, no associated vorticity. As the background plasma flow becomes more non-uniform in the radial direction, the curl of the velocity field now has components perpendicular to the direction of the flow. The background vorticity is localised to the interior of the magnetic cylinder, where the non-uniform plasma flow is located. The centre column in Figure (\ref{Gaussian_flow_vorticity_comparisons}) shows the perturbed velocity field for the slow body kink mode in the presence of a non-uniform flow. The uniform case on the top row has two clear nodes as predicted by uniform theory. As the background plasma flow becomes more non-uniform, these nodes shift closer together and a resulting vortical motion becomes clearer around the centre of the waveguide, where the magnitude of the background flow is greatest. It can be seen in all plots that the boundary of the waveguide, plotted as a blue line, is unaffected in all cases of non-uniform flow. This result is expected from the eigenfunctions shown in Figure (\ref{gaussian_flow_eigenfunction_slowkink}) which are locally unchanged at the boundary for all background flow profiles. The right hand column of Figure (\ref{Gaussian_flow_vorticity_comparisons}) shows the resulting background plus perturbed vorticity field. It can be seen that as the non-uniformity of background plasma flow is increased (further down the Figure column) that the vorticity is spread out over the whole region of inhomogeneity. Vortical motions become more apparent with increasing non-uniform flow which may act as a driver for other forms of MHD waves. Figure (\ref{3D_Gaussian_flow_Suzana}) again displays the 3D representation of the 2D velocity field seen in Figure (\ref{Gaussian_flow_vorticity_comparisons}). Locations of the nodes in the slow body kink eigenfunctions can be seen in Figure (\ref{3D_Gaussian_flow_Suzana_uniformflow}) in the velocity field vector for the uniform flow case. However, as expected the locations of these nodes are pushed together when the background plasma flow is modelled as a Gaussian profile, and is transported up through the tube with the propagation of the wave seen in Figure (\ref{3D_Gaussian_flow_Suzana_W06}).

\section{Conclusions}\label{conclusions}
In this paper a previously developed numerical technique described in \citet{Skirvin2021} has been employed to obtain the eigenvalues for trapped MHD waves in possible cylindrical environments representing some cases observed in the solar atmosphere. The algorithm has been tested against well known analytical results in a simple uniform cylindrical geometry and a more complex scenario that takes into account background twist of the equilibrium magnetic field. For both case studies the correct eigenvalues were obtained compared to those found in previous analytical studies that derive and solve the corresponding dispersion relation not required in this work. The tool was then applied to investigate the properties of MHD waves in non-uniform magnetic cylinders modelled by an inhomogeneous equilibrium plasma density and also an inhomogeneous background plasma flow. When the equilibrium plasma density is modelled as a series of Gaussian profiles with varying widths the spatial eigenfunctions are changed under both photospheric and coronal conditions. Firstly, under photospheric conditions, slow surface waves experience a cut off when the background density is sufficiently non-uniform. For our studies, a width ($W \approx 0.9$) that corresponds to the internal density at the boundary being halfway between $\rho_{0i}$ and $\rho_{0e}$ is sufficient enough to absorb these modes into the slow continuum. Furthermore, in the thin tube limit the fundamental kink branch no longer tends to the kink speed but instead favours an averaged value between $c_k$ and $c_{kB}$ due to the discontinuous nature of the density profile at the boundary. Finally, as the non-uniformity is increased, the frequency of magnetoacoustic waves decreases such that the band of body modes is also absorbed into the slow continuum at larger inhomogeneities. Comparisons of the spatial eigenfunctions for different wave modes revealed that the fast axisymemtric modes are not affected by the radial equilibrium inhomogeneity. The fast non-axisymmetric (kink) modes however, experience an increase in the azimuthal displacement at the boundary as the equilibrium plasma density becomes more non-uniform. The internal spatial structure of the slow body sausage and kink modes is greatly affected. Similar to the results found in Paper 1, additional nodes and points of inflexion appear as the background plasma density is modelled with a profile that is increasingly non-uniform. In both cases for the slow body modes of a non-uniform photospheric cylinder, the local amplitude of the eigenfunctions at the boundary is unaffected. Under coronal conditions, similar behaviour is observed with regards to the eigenfunctions. The fundamental kink mode tends to an averaged value between $c_k$ and $c_{kB}$ in the thin tube limit as the background plasma density becomes more inhomogeneous. The slow body modes are absorbed into the slow continuum with increasing non-uniform equilibria and these modes are no longer trapped solutions. Comparison between the eigenfunctions for the fast body sausage and kink modes reveal similar results to the photospheric cylinder. The local maximum amplitude of perturbation for the fast body sausage mode increases with increasing non-uniform plasma density, although this is not a significant change. The boundary value of the azimuthal displacement perturbation for the fundamental kink mode increases as the background plasma density is modelled as a clear Gaussian profile. To aid understanding in observations, a visual representation of this effect was provided. It is shown that as the background plasma density is increased, the boundary shape of the fundamental kink mode becomes distorted, possibly due to the linear regime of the onset of Kelvin-Helmholtz instability \citep{ant2014}.

The second case study analysed in this work investigated the behaviour of magnetoacoustic MHD wave modes in a coronal magnetic cylinder with a non-uniform background plasma flow. The plasma flow was again modelled as a series of Gaussian profiles with differing widths and the amplitude was kept small as to avoid any affects of flow related instabilities. The inhomogeneous plasma flow affected the forward and backward propagating slow body modes more than any other wave mode. Similar to the results found in Paper 1, we have found that the non-uniform flow creates an asymmetry between the phase speeds of forward and backward propagating slow body modes. Furthermore, like the case study investigating a background inhomogeneous density, the spatial eigenfunctions for slow kink and sausage body modes are affected due to the background flow. The eigenfunctions do not exhibit any changes at the boundary, similar to the behaviour of slow body modes in a photospheric cylinder with a non-uniform density. This is because body modes, unlike fast surface modes, propagate throughout the internal structure of the waveguide and not just amplified at the boundary. The background plasma flow introduces extra nodes into the spatial eigenfunctions at sufficient inhomogeneity and also changes the location of the local maximum in the spatial eigenfunction. Further investigation of vorticity due to the presence of a non-uniform background flow reveals that as the inhomogeneity of the background flow is increased, the resulting vorticity associated with the velocity perturbation also increases. It has been shown in this work that while the background vorticity increases with increasing equilibrium non-uniformity, the perturbed vorticity also increases. This is an important note to realise because it suggests that a non-uniform flow can produce a rotational perturbation. This motion may act as a driver to excite other forms of MHD waves e.g. Alfv\'{e}n waves. Therefore MHD modes in an inhomogeneous equilibrium can possibly self excite other MHD wave modes within the solar atmosphere.

The tool introduced in \citet{Skirvin2021} and further applied in this work has endless possible applications. Future work can extend that presented in this study to consider the leaky regime, which will provide information about wave damping and other wave phenomena including resonant absorption. Modelling the waveguide as a magnetic cylinder also allows a greater choice in the structure of the equilibrium plasma. Realistic models can be investigated that include individual investigations or combinations of radially non-uniform plasma, magnetic twist, rotational flow, non-linear equilibria etc.

The numerical code, Sheffield Dispersion Diagram Code (SDDC) introduced and applied in this work is available on the Plasma Dynamics Group (PDG) website \footnote{\url{https://sites.google.com/sheffield.ac.uk/pdg/solar-codes?authuser=0}} along with the user manual which explains some cases shown in this work. This code and the accompanying tools have been developed using Python an open-source and community-developed programming language.

\section*{Acknowledgements}
This work has been supported by STFC (UK). SJS is grateful to STFC for the PhD studentship project reference (2135820). VF and GV are grateful to Science and Technology Facilities Council (STFC) grant ST/V000977/1, and The Royal Society, International Exchanges Scheme, collaboration with Brazil (IES191114) and Chile (IE170301). SJS and VF would like to thank the International Space Science Institute (ISSI) in Bern, Switzerland, for the hospitality provided to the members
of the team on ‘The Nature and Physics of Vortex Flows in Solar Plasmas’. SJS and GV wish to acknowledge scientific discussions with the Waves in the Lower Solar Atmosphere (WaLSA; https://www.WaLSA.team) team, which is supported by the Research Council of Norway (project no. 262622). This research has also received financial support from the European Union’s Horizon 2020 research
and innovation program under grant agreement No. 824135 (SOLARNET). The authors thank M. Goossens for his comments that have greatly improved the quality of this manuscript.

\section*{Data Availability}
Accessibility of data used in this research is available upon request from the authors.



\bibliographystyle{mnras}
\bibliography{ref} 




\appendix

\section{Alfv\'{e}n and slow continuum in a non-uniform density cylinder }

\begin{figure*}
   \centering
   \begin{subfigure}{.49\textwidth}
        \centering
        \includegraphics[width=9.cm]{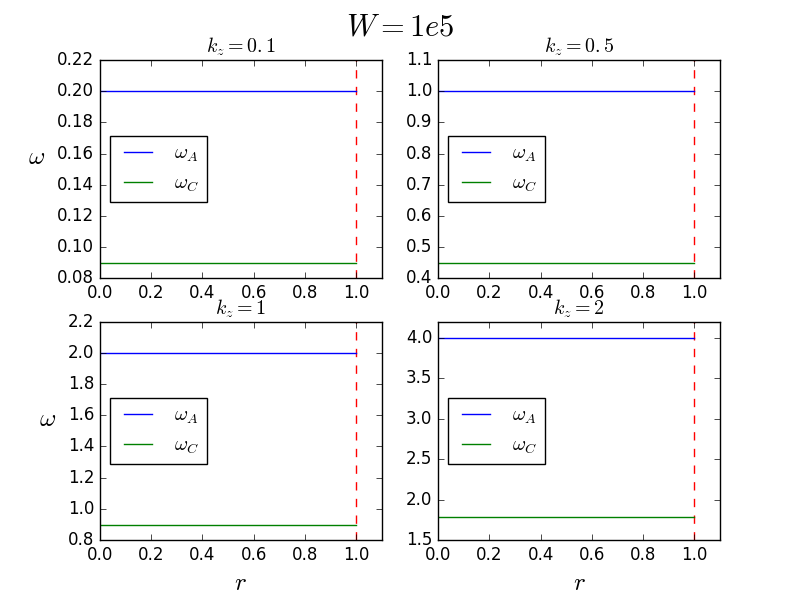}
        \caption{}
        \label{}
    \end{subfigure}
   \begin{subfigure}{.49\textwidth}
        \centering
        \includegraphics[width=9.cm]{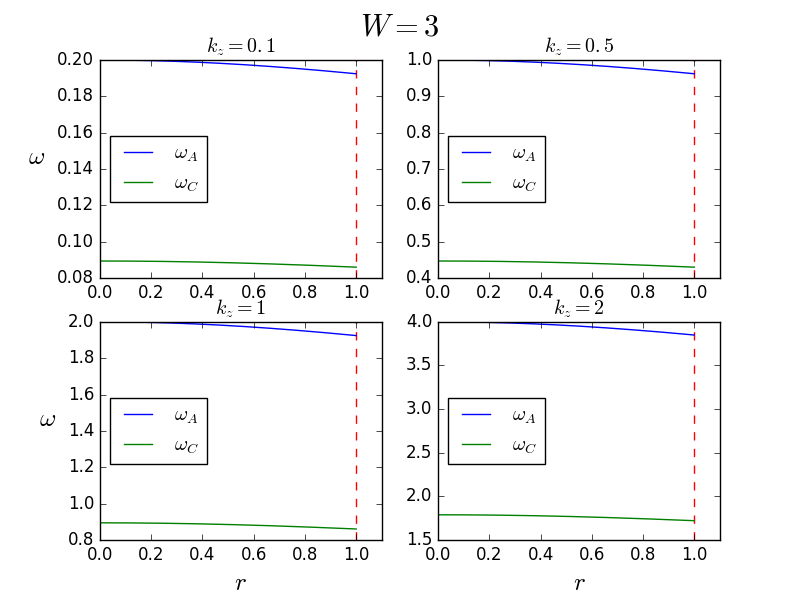}
        \caption{}
        \label{}
    \end{subfigure}   
   \begin{subfigure}{.49\textwidth}
        \centering
        \includegraphics[width=9.cm]{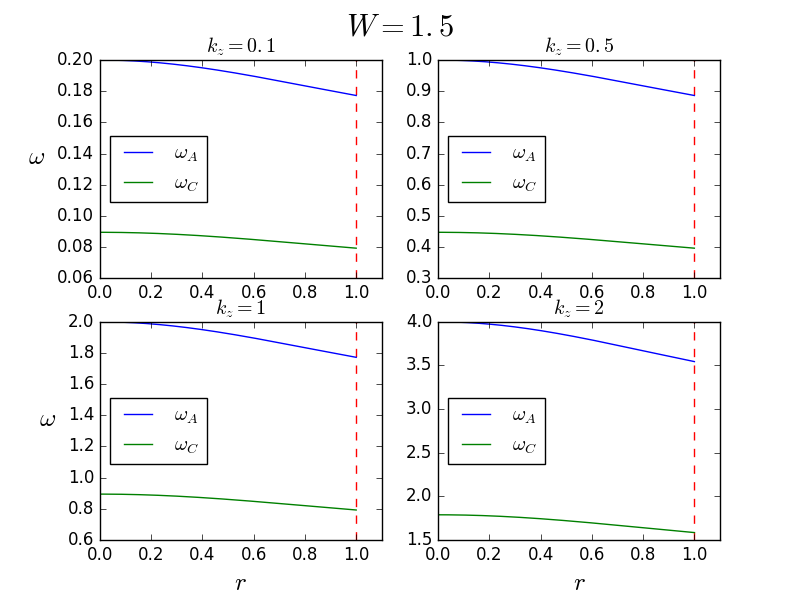}
        \caption{}
        \label{}
    \end{subfigure} 
   \begin{subfigure}{.49\textwidth}
        \centering
        \includegraphics[width=9.cm]{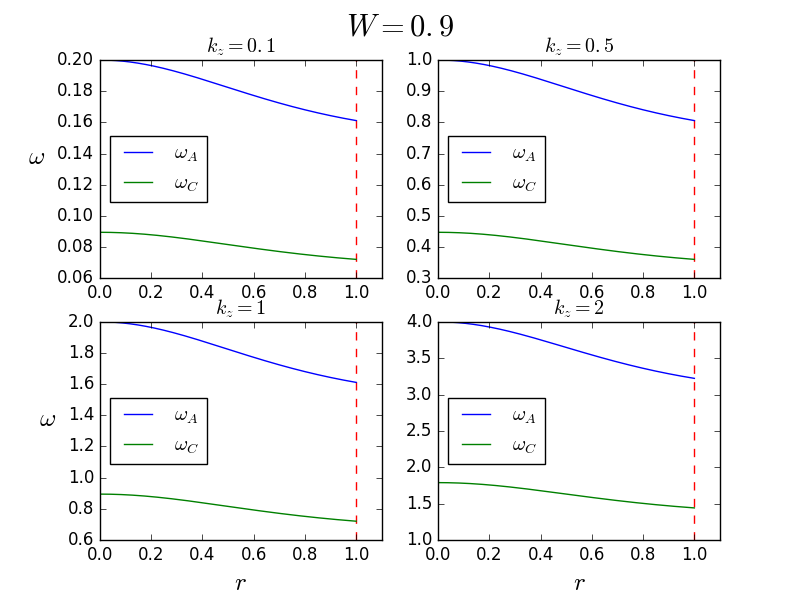}
        \caption{}
        \label{}
    \end{subfigure} 
    \caption{The Alfv\'{e}n (blue line) and slow continua (green line) are shown as a function of spatial variable $r$ for a non-uniform cylinder under photospheric conditions. For frequencies lying inside this range, the discrete wave modes are swallowed by the continua. These continua are shown for different wavenumber $k$. (a) The continua for the uniform density case, (b) $W = 3$, (c) $W = 1.5$, (d) $W = 0.9$.}
    \label{Photospheric_density_appendix}
   \end{figure*}

\begin{figure*}
   \centering
   \begin{subfigure}{.49\textwidth}
        \centering
        \includegraphics[width=9.cm]{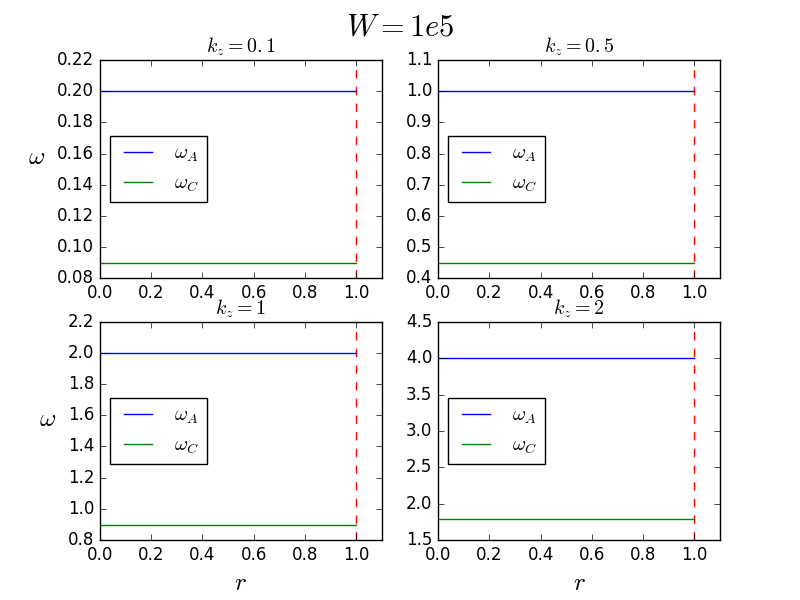}
        \caption{}
        \label{}
    \end{subfigure}
   \begin{subfigure}{.49\textwidth}
        \centering
        \includegraphics[width=9.cm]{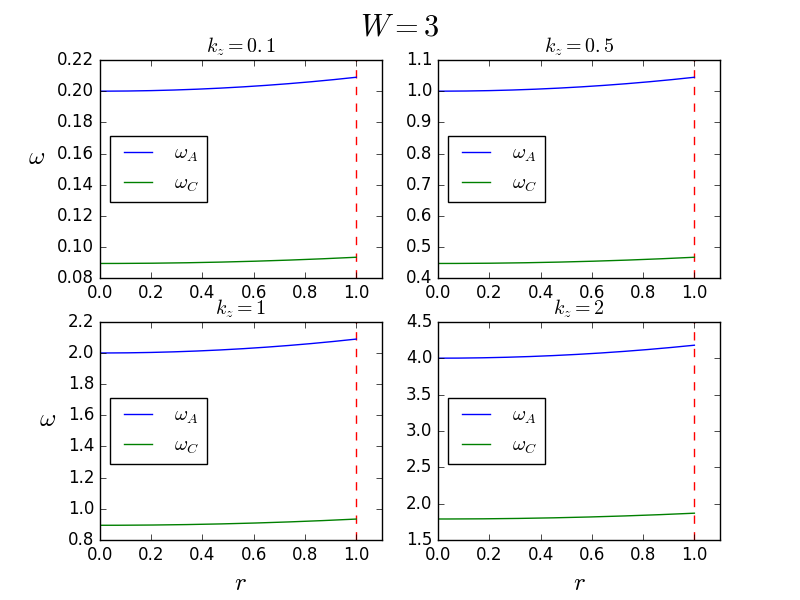}
        \caption{}
        \label{}
    \end{subfigure}   
   \begin{subfigure}{.49\textwidth}
        \centering
        \includegraphics[width=9.cm]{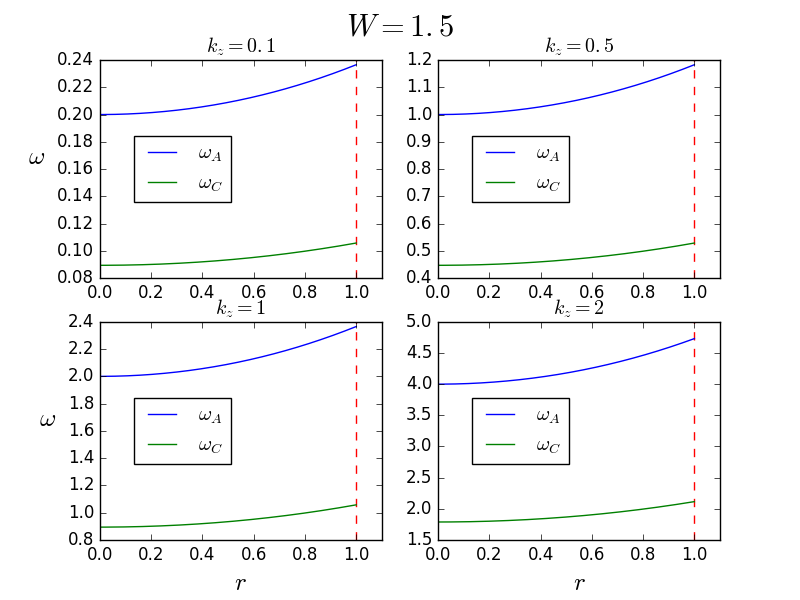}
        \caption{}
        \label{}
    \end{subfigure} 
   \begin{subfigure}{.49\textwidth}
        \centering
        \includegraphics[width=9.cm]{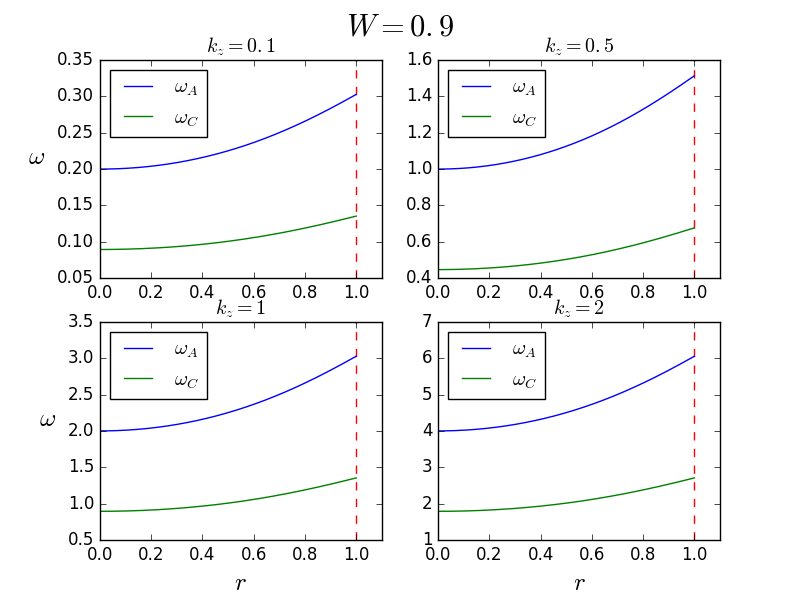}
        \caption{}
        \label{}
    \end{subfigure} 
    \caption{Same as Figure~\ref{Photospheric_density_appendix} but for the case of a non-uniform density cylinder under coronal conditions.}
    \label{coronal_density_appendix}
   \end{figure*}

The under-dense photospheric flux tube with a non-uniform internal density profile discussed in Section~\ref{nonuniform_photospheric_density} has characteristic frequencies $\omega_A$ and $\omega_c$ which depend on spatial variable $r$. Here, it may be instructive to plot the characteristic frequencies as a function of $r$ for different values of wavenumber $k$, in which it is linear. These plots are shown in Figure~\ref{Photospheric_density_appendix} for the photospheric case and Figure~\ref{coronal_density_appendix} for the coronal case considered in Section~\ref{Coronal_plasma_density_section}.

\bsp	
\label{lastpage}
\end{document}